\documentclass[12pt,titlepage,letterpaper]{utarticle}

\usepackage{amsmath}
\usepackage{mathrsfs}
\usepackage{amsfonts}
\usepackage{amssymb}
\usepackage{amsthm}
\usepackage{mathtools}
\usepackage{mathbbol}
\usepackage{graphicx}
\usepackage{color}
\usepackage{ucs}
\usepackage{xparse}
\usepackage{tikz}
\usepackage{tocloft}
\usepackage[titletoc,title]{appendix}
\usepackage[nosort]{cite}
\usepackage{hyperref}
\usepackage{longtable}

\usepackage{caption}

\definecolor{aqua}{rgb}{0, 1.0, 1.0}
\definecolor{fuschia}{rgb}{1.0, 0, 1.0}
\definecolor{gray}{rgb}{0.502, 0.502, 0.502}
\definecolor{lime}{rgb}{0, 1.0, 0}
\definecolor{maroon}{rgb}{0.502, 0, 0}
\definecolor{navy}{rgb}{0, 0, 0.502}
\definecolor{olive}{rgb}{0.502, 0.502, 0}
\definecolor{purple}{rgb}{0.502, 0, 0.502}
\definecolor{silver}{rgb}{0.753, 0.753, 0.753}
\definecolor{teal}{rgb}{0, 0.502, 0.502}
\definecolor{darkgreen}{rgb}{0,0.502,0}

\makeatletter
\newdimen\itex@wd%
\newdimen\itex@dp%
\newdimen\itex@thd%
\def\itexspace#1#2#3{\itex@wd=#3em%
\itex@wd=0.1\itex@wd%
\itex@dp=#2ex%
\itex@dp=0.1\itex@dp%
\itex@thd=#1ex%
\itex@thd=0.1\itex@thd%
\advance\itex@thd\the\itex@dp%
\makebox[\the\itex@wd]{\rule[-\the\itex@dp]{0cm}{\the\itex@thd}}}
\makeatother

\makeatletter
\newif\if@sup
\newtoks\@sups
\def\append@sup#1{\edef\act{\noexpand\@sups={\the\@sups #1}}\act}%
\def\reset@sup{\@supfalse\@sups={}}%
\def\mk@scripts#1#2{\if #2/ \if@sup ^{\the\@sups}\fi \else%
  \ifx #1_ \if@sup ^{\the\@sups}\reset@sup \fi {}_{#2}%
  \else \append@sup#2 \@suptrue \fi%
  \expandafter\mk@scripts\fi}
\def\tensor#1#2{\reset@sup#1\mk@scripts#2_/}
\def\multiscripts#1#2#3{\reset@sup{}\mk@scripts#1_/#2%
  \reset@sup\mk@scripts#3_/}
\makeatother

\makeatletter
\newbox\slashbox \setbox\slashbox=\hbox{$/$}
\def\itex@pslash#1{\setbox\@tempboxa=\hbox{$#1$}
  \@tempdima=0.5\wd\slashbox \advance\@tempdima 0.5\wd\@tempboxa
  \copy\slashbox \kern-\@tempdima \box\@tempboxa}
\def\slash{\protect\itex@pslash}
\makeatother

\def\clap#1{\hbox to 0pt{\hss#1\hss}}

\def\mathclap{\mathpalette\mathclapinternal}

\def\mathclapinternal#1#2{\clap{$\mathsurround=0pt#1{#2}$}}

\let\oldroot\root
\def\root#1#2{\oldroot #1 \of{#2}}
\renewcommand{\sqrt}[2][]{\oldroot #1 \of{#2}}

\DeclareSymbolFont{symbolsC}{U}{txsyc}{m}{n}
\SetSymbolFont{symbolsC}{bold}{U}{txsyc}{bx}{n}
\DeclareFontSubstitution{U}{txsyc}{m}{n}

\DeclareSymbolFont{stmry}{U}{stmry}{m}{n}
\SetSymbolFont{stmry}{bold}{U}{stmry}{b}{n}

\DeclareFontFamily{OMX}{MnSymbolE}{}
\DeclareSymbolFont{mnomx}{OMX}{MnSymbolE}{m}{n}
\SetSymbolFont{mnomx}{bold}{OMX}{MnSymbolE}{b}{n}
\DeclareFontShape{OMX}{MnSymbolE}{m}{n}{
    <-6>  MnSymbolE5
   <6-7>  MnSymbolE6
   <7-8>  MnSymbolE7
   <8-9>  MnSymbolE8
   <9-10> MnSymbolE9
  <10-12> MnSymbolE10
  <12->   MnSymbolE12}{}

\makeatletter
\def\re@DeclareMathSymbol#1#2#3#4{%
    \let#1=\undefined
    \DeclareMathSymbol{#1}{#2}{#3}{#4}}
\re@DeclareMathSymbol{\neArrow}{\mathrel}{symbolsC}{116}
\re@DeclareMathSymbol{\neArr}{\mathrel}{symbolsC}{116}
\re@DeclareMathSymbol{\seArrow}{\mathrel}{symbolsC}{117}
\re@DeclareMathSymbol{\seArr}{\mathrel}{symbolsC}{117}
\re@DeclareMathSymbol{\nwArrow}{\mathrel}{symbolsC}{118}
\re@DeclareMathSymbol{\nwArr}{\mathrel}{symbolsC}{118}
\re@DeclareMathSymbol{\swArrow}{\mathrel}{symbolsC}{119}
\re@DeclareMathSymbol{\swArr}{\mathrel}{symbolsC}{119}
\re@DeclareMathSymbol{\nequiv}{\mathrel}{symbolsC}{46}
\re@DeclareMathSymbol{\Perp}{\mathrel}{symbolsC}{121}
\re@DeclareMathSymbol{\Vbar}{\mathrel}{symbolsC}{121}
\re@DeclareMathSymbol{\sslash}{\mathrel}{stmry}{12}
\re@DeclareMathSymbol{\boxslash}{\mathrel}{stmry}{27}
\re@DeclareMathSymbol{\boxbslash}{\mathrel}{stmry}{28}
\re@DeclareMathSymbol{\boxast}{\mathrel}{stmry}{24}
\re@DeclareMathSymbol{\boxcircle}{\mathrel}{stmry}{29}
\re@DeclareMathSymbol{\boxbox}{\mathrel}{stmry}{30}
\re@DeclareMathSymbol{\obslash}{\mathrel}{stmry}{20}
\re@DeclareMathSymbol{\obar}{\mathrel}{stmry}{58}
\re@DeclareMathSymbol{\olessthan}{\mathrel}{stmry}{60}
\re@DeclareMathSymbol{\ogreaterthan}{\mathrel}{stmry}{61}
\re@DeclareMathSymbol{\bigsqcap}{\mathop}{stmry}{"64}
\re@DeclareMathSymbol{\biginterleave}{\mathop}{stmry}{"6}
\re@DeclareMathSymbol{\invamp}{\mathrel}{symbolsC}{77}
\re@DeclareMathSymbol{\parr}{\mathrel}{symbolsC}{77}
\makeatother

\makeatletter
\def\Decl@Mn@Delim#1#2#3#4{%
  \if\relax\noexpand#1%
    \let#1\undefined
  \fi
  \DeclareMathDelimiter{#1}{#2}{#3}{#4}{#3}{#4}}
\def\Decl@Mn@Open#1#2#3{\Decl@Mn@Delim{#1}{\mathopen}{#2}{#3}}
\def\Decl@Mn@Close#1#2#3{\Decl@Mn@Delim{#1}{\mathclose}{#2}{#3}}
\Decl@Mn@Open{\llangle}{mnomx}{'164}
\Decl@Mn@Close{\rrangle}{mnomx}{'171}
\Decl@Mn@Open{\lmoustache}{mnomx}{'245}
\Decl@Mn@Close{\rmoustache}{mnomx}{'244}
\makeatother

\makeatletter
\DeclareRobustCommand\widecheck[1]{{\mathpalette\@widecheck{#1}}}
\def\@widecheck#1#2{%
    \setbox\z@\hbox{\m@th$#1#2$}%
    \setbox\tw@\hbox{\m@th$#1%
       \widehat{%
          \vrule\@width\z@\@height\ht\z@
          \vrule\@height\z@\@width\wd\z@}$}%
    \dp\tw@-\ht\z@
    \@tempdima\ht\z@ \advance\@tempdima2\ht\tw@ \divide\@tempdima\thr@@
    \setbox\tw@\hbox{%
       \raise\@tempdima\hbox{\scalebox{1}[-1]{\lower\@tempdima\box
\tw@}}}%
    {\ooalign{\box\tw@ \cr \box\z@}}}
\makeatother

\makeatletter
\NewDocumentCommand\mathraisebox{moom}{%
\IfNoValueTF{#2}{\def\@temp##1##2{\raisebox{#1}{$\m@th##1##2$}}}{%
\IfNoValueTF{#3}{\def\@temp##1##2{\raisebox{#1}[#2]{$\m@th##1##2$}}%
}{\def\@temp##1##2{\raisebox{#1}[#2][#3]{$\m@th##1##2$}}}}%
\mathpalette\@temp{#4}}
\makeatletter

\makeatletter
\def\udots{\mathinner{\mkern2mu\raise\p@\hbox{.}
\mkern2mu\raise4\p@\hbox{.}\mkern1mu
\raise7\p@\vbox{\kern7\p@\hbox{.}}\mkern1mu}}
\makeatother

\theoremstyle{plain}

\theoremstyle{definition}

\theoremstyle{remark}

\begin{document}

\preprint{
UTTG--28--2021\\
}

\title{Nonabelian Twists of the $D_4$ Theory}

\author{Jacques Distler
   \address{
       Theory Group\\
      Department of Physics,\\
      University of Texas at Austin\\
      Austin, TX 78712, USA \\
      {~}\\
      \email{distler@golem.ph.utexas.edu}\\
      \email{ali.shehper1@gmail.com}
  },
  Behzat Ergun
  \address{
    Department of Physics,\\
    Technion\\
    Israel Institute of Technology\\
    Haifa, 32000, Israel\\
    {~}\\
    \email{bergun@utexas.edu}
  } and Ali Shehper ${}^\mathrm{a}$
}
\date{\today}

\Abstract{
We study theories of type $D_4$ in class-S, with nonabelian outer-automorphism twists around various cycles of the punctured Riemann surface $C$. We propose an extension of previous formul\ae\ for the superconformal index to cover this case and classify the SCFTs corresponding to fixtures (3-punctured spheres). We then go on to study families of SCFTs corresponding to once-punctured tori and 4-punctured spheres. These exhibit new behaviours, not seen in previous investigations. In particular, the generic theory with 4 punctures on the sphere from non-commuting $\mathbb{Z}_2$ twisted sectors has six distinct weakly-coupled descriptions.
}

\maketitle

\tocloftpagestyle{empty}
\tableofcontents
\vfill
\newpage
\setcounter{page}{1}

\section{Introduction}\label{introduction}
Class $S$ theories provide a vast class of examples of four-dimensional $\mathcal{N}=2$ superconformal field theories obtained by partially-twisted compactifications of the six-dimensional $(2,0)$ theory of type $\mathfrak{j} = A, D, E$ on a possibly punctured Riemann surface \cite{Gaiotto:2009we,Gaiotto:2009hg}.
The exactly marginal deformations of the four-dimensional theory are parameterized by the moduli space of complex structures $\overline{\mathcal{M}}_{g,n}$ or a branched cover of it. At the intersection of various boundary divisors of $\overline{\mathcal{M}}_{g,n}$, the Riemann surface develops nodal curves which are 3-punctured spheres. In such a degeneration limit, the SCFT is realized as a product of some free vector multiplets and the SCFTs corresponding to the 3-punctured spheres. Moving to different points of intersection along the boundary, we get different decompositions  and the various decompositions are said to be S-dual to each other. Thus, one can study systematically the building blocks of this procedure, i.e. the 3-punctured spheres, and the possible ways to gauge them in order to study SCFTs associated to arbitrary Riemann surfaces. 

It is also possible to allow twists of the outer-automorphism of the Lie algebra $\mathfrak{j}$ along various cycles of the surface \cite{Tachikawa:2009rb,Tachikawa:2010vg}. This significantly enlarges the zoo of SCFTs obtained from the six-dimensional origin. A complete classification of the three-punctured spheres in almost all of the sectors with or without the outer automorphism twists has been carried out (see   
\cite{Chacaltana:2017boe,Chacaltana:2018zag,Beem:2020pry} and the references therein).  The only remaining case is the twisted $D_4$ sector where the twists are allowed to be in the full nonabelian outer automorphism group $\text{Out}(D_4)=S_3$ --- the symmetric group on three letters.\footnote{Some properties of SCFTs in this sector are known. For example, the authors of \cite{Bhardwaj:2021pfz,Bhardwaj:2021ojs,Bhardwaj:2021wif} proposed methods to calculate 1-form symmetry, the global form of 0-form symmetry, and 2-group symmetry groups of all theories of class $S$ including those of $D_4$ sector with nonabelian twists.} The nonabelian nature of twists introduces several new challenges which we address in this paper.

If the twists are constrained to lie in an abelian subgroup of $S_3$, i.e. a $\mathbb{Z}_2$ or a $\mathbb{Z}_3$ subgroup, the classification of three-punctured spheres has already been carried out in \cite{Chacaltana:2013oka} and \cite{Chacaltana:2016shw} respectively. Thus the local data associated to punctures is already known. The main difficulty appears in understanding the cutting and gluing of surfaces.\footnote{In the case of $\mathbb{Z}_2$ or $\mathbb{Z}_3$ twists, the classification of surfaces was relatively straightforward. The twists on a surface $C$ are classified by $H^1(C-\{p_i\},\mathbb{Z}_2)$ and $H^1(C-\{p_i\},\mathbb{Z}_3)$ respectively (where $p_i$ denotes the location of $i^{\text{th}}$ puncture), and there is a Mayer-Vietoris principle for these groups that helps understand the cutting and gluing of surfaces decorated with twists.}  In particular, 
\begin{itemize}
\item We are forced to study $S_3$-bundles on a surface $C$, which are classified by elements of $\text{Hom}(\pi_1(C - \{p_i\}),S_3)$ up to conjugation. How to consistently glue these surfaces so that we can build up $S_3$-bundles on arbitrary surfaces presents a challenge.

\item Even at the level of three-punctured spheres, a new sector appears where the twists around the punctures are in non-commuting sectors. The superconformal index \cite{Kinney:2005ej,Gadde:2009kb,Gadde:2011ik,Gadde:2011uv}, which has been studied in \cite{Lemos:2012ph,Chacaltana:2016shw} in the presence of abelian twists of $D$-type theories and is an invaluable tool for the purpose of classification, is not previously known for this sector. 
\end{itemize}

In this paper we solve these issues. As a consequence, we find new features which were previously absent in the study of class $S$ (and in fact, in the study of 4d $\mathcal{N}=2$ SCFTs in general). The most interesting of these is the presence of new one-dimensional conformal manifolds parameterizing the exactly marginal deformations of SCFTs associated to four-punctured spheres. We find, contrary to all of the previously known examples, that these conformal manifolds can have more than three and up to six different weakly coupled limits.\footnote{All previously known one-dimensional $\mathcal{N}=2$ conformal manifolds are known to have at most 3 distinct weakly coupled limits. This is true for the conformal manifolds that parametrize the space of exactly marginal couplings of Lagrangian field theories as well as the ones that have gaugings of isolated interacting SCFTs as their only weakly coupled limits. A familiar example of the former case is $SU(N)$ gauge theory with $2N$ hypermultiplets. The conformal manifold has $1$ or $2$ distinct weakly coupled limits depending on whether $N=2$ or $N\geq 3$ respectively \cite{Seiberg:1994aj,Argyres:1995wt,Argyres:2007cn,Gaiotto:2009we}.} We will study these conformal manifolds in more detail in the upcoming paper \cite{Distler:toappear}, but here we already give two examples in \S\ref{4punctured_sphere_with_4_twisted_punctures}.

This paper is organized as follows. In \S\ref{nonabelian_twists_of_the__theory_and__bundles}, we provide a description of $S_3$ bundles on three-punctured spheres that is suitable to cutting and gluing in the class $S$ program. In \S\ref{superconformal_index_for_nonabelian_twisted__theories}, we give superconformal indices of fixtures in various nonabelian twisted $D_4$ sectors. \S\ref{oncepunctured_torus} contains a discussion of once-punctured tori in these sectors. In \S\ref{classification_of_fixtures}, we present a classification of fixtures in the new sector that we mentioned above. We turn to 4-punctured spheres in \S\ref{4punctured_sphere_with_4_twisted_punctures}, presenting the general setup for 4 punctures in  (non-commuting) $\mathbb{Z}_2$-twisted sectors. In \S\ref{resolving_atypical_punctures}, we study atypical punctures in the $\mathbb{Z}_3$-twisted sector which resolve to a pair of punctures in non-commuting $\mathbb{Z}_2$-twisted sectors. In
\S\ref{six_weak}, we give an example of a family of $\mathcal{N}=2$ SCFTs with 6 distinct weak-coupling limits, whose conformal manifold is (the compactification of) $\text{UHP}/\Gamma(4)$. More details, and the general story for other sectors, will be discussed in \cite{Distler:toappear}.

\section{Nonabelian twists of the $D_4$ theory and $S_3$ bundles}\label{nonabelian_twists_of_the__theory_and__bundles}

The twisted compactifications of the $(2,0)$ theory of type-$\mathfrak{j}$ on a (punctured) Riemann surface, $C$, involve the following data.

\begin{itemize}%
\item A choice of principal $\Gamma$-bundle, $P$, on $C$, where $\Gamma\subset \text{Out}(\mathfrak{j})$. Let $\gamma_i\in\Gamma$ be the group element singled out by restricting $P$ to the circle surrounding the $i^{\text{th}}$ puncture (with the choice of a basepoint).
\item For each $i$, a choice of nilpotent orbit in $\mathfrak{g}$, the Langlands dual of $\mathfrak{g}^\vee\subset \mathfrak{j}$, the subalgebra invariant under the action of $\gamma_i$.
\end{itemize}

Note that above, we required the choice of an actual bundle on $C$, and not just its isomorphism class. This is because in the tinkertoy program, we build up more complicated surfaces, $C$, by gluing together 3-punctured spheres (``fixtures''). In order to glue, we need to glue actual bundles, rather than isomorphism classes thereof. More precisely, we need (at least) an actual bundle on each boundary circle (since these circles are the loci along which we glue)\footnote{In the examples we study in this paper  (the once-punctured torus and the four-punctured sphere), the isomorphism class of the bundle on the sewn surface is simply related to the data on the 3-punctured sphere(s) that we sew together along a single circle. We can always choose the basepoint (in our description of isomorphism classes of bundles) to lie on that circle. (For the 4-punctured sphere, there are three distinct ways to decompose the surface into 3-punctured spheres and we can choose the three circles to intersect at the basepoint.) For higher genus or more punctures, we necessarily have to sew along disjoint circles so in  \cite{Distler:toappear} we really will find the more complicated formalism, that we introduce here, useful.}.

To this end, we choose a CW-decomposition of the 3-punctured sphere. For the $0$-skeleton, pick a point on each of the three boundaries of the 3-punctured sphere. Let the six 1-cells be the three boundary circles (connecting each 0-cell with itself) and three arcs connecting pairs of 0-cells. The 2-cells are then a triangle and a hexagon.

\begin{equation} 
\begin{matrix}
 \includegraphics[width=157pt]{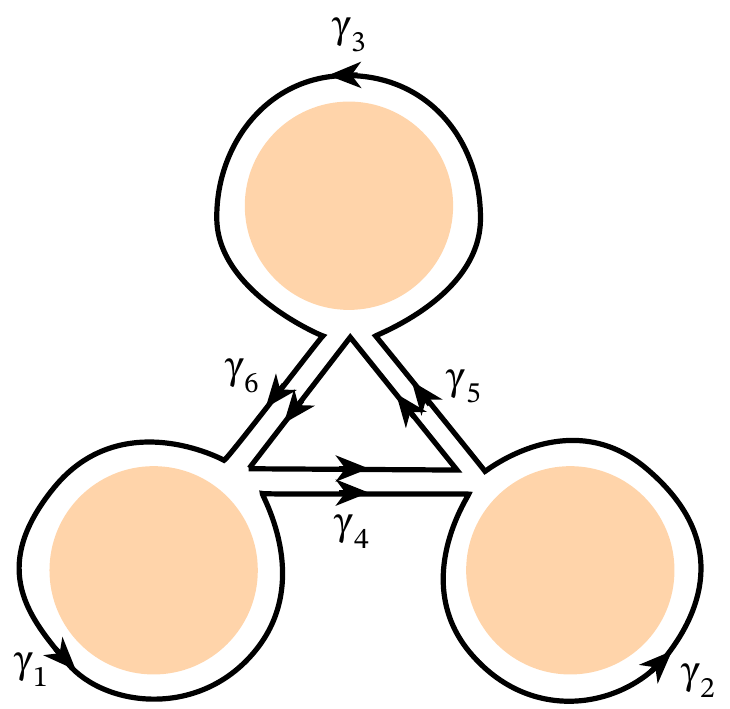}
 \end{matrix}
\label{unzipped}\end{equation}
$P$ can be trivialized over each 0-cell. We then assign an element of $\Gamma$ to each 1-cell, which tells us how the two trivializations at the endpoints compare. (N.b.: the choice of $\gamma\in\Gamma$ associated to a 1-cell depends on the orientation; if we reverse the orientation of the 1-cell, $\gamma$ is replaced by $\gamma^{-1}$.) Physicists can think of this as the holonomy of the (unique and flat, since $\Gamma$ is discrete) connection on $P$.

The $\Gamma$-bundle on each boundary circle is uniquely specified by its holonomy. We will refer to these three group elements, $\gamma_1,\gamma_2,\gamma_3$, as the ``external twists''. They determine the restriction of $P$ to a principal $\Gamma$-bundle on each boundary circle. The group elements, $\gamma_4, \gamma_5, \gamma_6$, associated to the open arcs obey two compatibility conditions. Since the boundary of each 2-cell is a contractible path on $C$, the holonomy around it must be $\mathbb{1}$.

\begin{equation}
\begin{split}
\gamma_6\gamma_5\gamma_4&=\mathbb{1}\\
\gamma_6\gamma_3\gamma_5\gamma_2\gamma_4\gamma_1&=\mathbb{1}
\end{split}
\label{relations}\end{equation}
When $\Gamma$ is abelian, this simplifies to

\begin{displaymath}
\begin{split}
\gamma_6\gamma_5\gamma_4&=\mathbb{1}\\
\gamma_3\gamma_2\gamma_1&=\mathbb{1}
\end{split}
\end{displaymath}
We will refer to $\gamma_4, \gamma_5, \gamma_6$ as ``internal twists''.

We shall see in \S\ref{superconformal_index_for_nonabelian_twisted__theories} that the physics of a fixture may depend on this more refined data (i.e. the choice of $\Gamma$-bundles on three boundary circles, and the data of how these $\Gamma$-bundles compare). In particular, when the restriction of $P$ on all boundary circles is trivial $(\gamma_1 = \gamma_2 = \gamma_3 = \mathbb{1})$, the only principle $\Gamma$-bundle is the trivial bundle. However, the physics depends on the internal twists $(\gamma_4, \gamma_5, \gamma_6)$ defined above (cf. \S\ref{_sector_5}). 

It is also useful to adopt this parametrization as the internal twists play a crucial role in determining the $\Gamma$-bundles that result when we start sewing together 3-punctured spheres to form more complicated Riemann surfaces.
As a simple example, consider $A_{2N-1}$ theory with one simple puncture (at $p_1$) and two full punctures (at $p_2$ and $p_3$). In this case, $\Gamma=\mathbb{Z}_2$ (which we denote multiplicatively). When $\gamma_1=\gamma_2=\gamma_3=1$, the fixture is $(2N)^2$ free hypermultiplets for any choice of $\gamma_{4,5,6}$. But now consider sewing together punctures 2,3 to form a once-punctured torus. When $\gamma_5=1$, this is an $SU(2N)$ gauge theory with hypermultiplets in the $(2N)\otimes \overline{(2N)}=\text{adjoint}\oplus (1)$. When $\gamma_5=-1$, the hypermultiplets transform as the $(2N)\otimes (2N) =(\begin{matrix} \includegraphics[width=17pt]{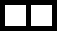}\end{matrix})\oplus
\left(\begin{matrix} \includegraphics[width=8pt]{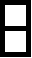}\end{matrix}\right)$.

Different choices of data defined in \eqref{unzipped} may be related by some simple operations. For example, conjugation
\begin{equation}
C_{h_1,h_2,h_3}\colon (\gamma_1,\gamma_2,\gamma_3,\gamma_4,\gamma_5,\gamma_6) \mapsto 
(h_1 \gamma_1 h_1^{-1}, h_2 \gamma_2 h_2^{-1}, h_3 \gamma_3 h_3^{-1}, h_2\gamma_4 h_1^{-1}, h_3\gamma_5 h_2^{-1}, h_1\gamma_6 h_3^{-1})
\label{conjugation}\end{equation}
for $h_i \in \Gamma$ yields, in general, a different solution to \eqref{relations}. The set of all conjugations is a group under composition, and it acts transitively on the set of all solutions to \eqref{relations} with fixed $([\gamma_1], [\gamma_2], [\gamma_3])$. We will see that this action is useful in understanding the physics of fixtures with the same $([\gamma_1], [\gamma_2], [\gamma_3])$ but different values of external and internal twists.

\subsection{$\Gamma=S_3$}\label{dfwwvcvv}

Let us turn now to the case of interest $\Gamma=S_3$,

\begin{displaymath}
\Gamma=\langle\alpha,\beta|\alpha^2=(\alpha\beta)^2=\beta^3=\mathbb{1}\rangle.
\end{displaymath}
For later convenience, let us tabulate the solutions to \eqref{relations}. There are two selection rules on the external twists $(\gamma_1,\gamma_2,\gamma_3)$:

\begin{itemize}%
\item If two of the external twists are trivial, then so is the third one.
\item The total parity must be be even\begin{displaymath}
\sigma(\gamma_1)\sigma(\gamma_2)\sigma(\gamma_3)=1
\end{displaymath}
where $\sigma$ is the sign representation ($\sigma(\beta)=1,\sigma(\alpha)=-1$).

\end{itemize}
For brevity, we will order the punctures so that $\sigma(\gamma_1)=1$ and use the conjugation \eqref{conjugation} to pick a representative of each conjugacy class for the external twists

\begin{longtable}{|l|l|}
\hline
$(\gamma_1,\gamma_2,\gamma_3)$&Solutions to \eqref{relations}\\
\endhead
\hline 
$(\mathbb{1},\mathbb{1},\mathbb{1})$&$\gamma_4=\text{any},\ \gamma_5=\text{any}$\\
\hline 
$(\mathbb{1},\alpha,\alpha)$&$\gamma_4=\text{any},\ \gamma_5=\mathbb{1},\alpha$\\
\hline 
$(\mathbb{1},\beta,\beta^2)$&$\gamma_4=\text{any},\ \gamma_5=\beta^l$\\
\hline 
$(\beta,\alpha,\alpha\beta)$&$\begin{aligned}\gamma_4&=\beta^j,& \ \gamma_5&=\mathbb{1},\alpha\beta\\ \gamma_4&=\alpha\beta^j,&\  \gamma_5&=\alpha,\beta^2\end{aligned}$\\
\hline 
$(\beta,\beta,\beta)$&$\gamma_4=\beta^j ,\ \gamma_5=\beta^l$\\
\hline 
\end{longtable}
\noindent where, of course, $\gamma_6=\gamma_4^{-1}\gamma_5^{-1}$. These are all of the cases where $(\gamma_4,\gamma_5,\gamma_6)=(\mathbb{1},\mathbb{1},\mathbb{1})$ is contained in the set of solutions. All of the other cases may be obtained by cyclically permuting the punctures and/or applying \eqref{conjugation}. For instance,

\begin{longtable}{|l|l|}
\hline
$(\gamma_1,\gamma_2,\gamma_3)$&Solutions to \eqref{relations}\\
\hline 
$(\mathbb{1},\alpha,\alpha\beta)$&$\gamma_4=\text{any},\ \gamma_5=\beta,\alpha\beta^2$\\
\hline 
$(\beta,\alpha,\alpha)$&$\begin{aligned}\gamma_4&=\beta^j,&\gamma_5&=\beta^2,\alpha\beta^2\\ \gamma_4&=\alpha\beta^j,&\ \gamma_5&=\beta,\alpha\beta\end{aligned}$\\
\hline 
$(\beta,\beta,\beta^2)$&$\gamma_4=\beta^j,\ \gamma_5=\alpha\beta^l$\\
\hline 
\end{longtable}

\section{Superconformal index for nonabelian twisted $D_4$ theories}\label{superconformal_index_for_nonabelian_twisted__theories}

In this section, we propose formulas for the superconformal index of SCFTs associated to the three-punctured spheres in the presence of external and internal twists. Our general philosophy is that the conjugation operation $C_{h_1,h_2,h_3}$ in \eqref{conjugation} acts at the level of the superconformal index in a simple way, namely, it acts on the weights associated to the $i^{\text{th}}$ puncture by the outer-automorphism $h_i$. It follows that the action of conjugation on the twisted punctures is trivial as the Lie algebras $\mathfrak{sp}(3)$ and $\mathfrak{g}_2$ have trivial outer automorphism groups.

We will organize our formulae in terms of the conjugacy classes of the external twists, $([\gamma_1],[\gamma_2],[\gamma_3])$. The action of conjugation helps us determine the dependence of the superconformal index on actual representatives of the conjugacy classes $[\gamma_i]$ as well as any internal twists on the fixture. For example, an immediate consequence of our proposal is that the superconformal index of fixtures in the $([\beta],[\beta],[\beta])$ and $([\beta],[\alpha],[\alpha])$ sectors are completely independent of these choices.

\subsection{$([\beta],[\beta],[\beta])$ sector}\label{_sector}

The superconformal index in the $(\beta,\beta,\beta)$ sector was studied in \cite{Chacaltana:2016shw}. As mentioned above, the same formula holds if we replace any or all of the $\beta$-twisted punctures by $\beta^2$ and/or introduce any internal twists on the fixture.

\subsection{$([\beta], [\alpha], [\alpha])$ sector}\label{_sector_2}

The superconformal index of a fixture in the $([\beta], [\alpha], [\alpha])$ sector has not been studied previously in the literature. Here we will propose a formula and provide some consistency checks for it. The same formula holds for any fixture in this sector regardless of the choice of internal twists and/or actual representatives of conjugacy classes of external twists.

A fixture in the $([\beta], [\alpha], [\alpha])$ sector is labeled by a $\mathfrak{g}_2$ nilpotent orbit, $O_1$, and two $\mathfrak{sp}(3)$ nilpotent orbits $O_{2,3}$. For each of the punctures, the nilpotent orbit determines an embedding $\rho: \mathfrak{su}(2) \rightarrow \mathfrak{g}$ where $\mathfrak{g}^\vee$ is the invariant subalgebra of $\mathfrak{so}(8)$ under the outer automorphism around the puncture. The flavor symmetry $\mathfrak{f}$ for the puncture is given by the centralizer of the image of $\rho$ in $\mathfrak{g}$. We label the fugacities of the embedding $\mathfrak{su}(2)$ as $\tau$ and the fugacities associated to the flavor symmetry of the three punctures as $a$, $b$, and $c$ respectively.

We will focus on the Schur limit of the superconformal index in this paper because it is easier to compute and counts precisely the multiplets that we use to classify our fixtures. Any other limit of the superconformal index \cite{Gadde:2011uv} can be computed after replacing the characters with the corresponding polynomials and using the appropriate K-factors in the formula below. Our proposal for the Schur index is 

\begin{equation}\label{betaalphaalpha}
\begin{split}
I^{([\beta], [\alpha], [\alpha])}_{Schur} (\tau)& = \frac{K(a(O_1), \tau) K(b(O_2), \tau) K(c(O_3), \tau )}{K([7,1],\tau)} \\
&\times \sum_{(n_1, n_2)} \frac{\chi_{G_2}^{(n_1, n_2)} (a(O_1), \tau) \chi_{Sp(3)}^{(n_2, n_1, n_2)} (b(O_2), \tau)
 \chi_{Sp(3)}^{(n_2, n_1, n_2)}(c(O_3), \tau)}
{\chi_{SO(8)}^{(n_2, n_1, n_2, n_2)} ([7,1],\tau)}.
\end{split}
\end{equation}

Here, $(n_1,n_2)$, $(n_2,n_1,n_2)$ and $(n_2, n_1, n_2, n_2)$ are the Dynkin labels of finite dimensional representations of $\mathfrak{g}_2$, $\mathfrak{sp}(3)$ and $\mathfrak{so}(8)$ respectively. The polynomials $\chi^{(n_1,n_2)}_{G_2}$, $\chi^{(n_2,n_1,n_2)}_{Sp(3)}$ and $\chi_{SO(8)}^{(n_2, n_1, n_2, n_2)}$ are the characters for these representations with the corresponding Dynkin labels. The flavor fugacities $a(O)$ associated to the nilpotent orbit $O$ of the corresponding Lie algebra $\mathfrak{g}$ labeling a puncture are determined by decomposing the fundamental representation of $\mathfrak{g}$ as a representation of $\rho(\mathfrak{su}(2)) \oplus \mathfrak{f}$. The K-factor for each puncture is determined by decomposing the adjoint representation of $\mathfrak{g}$ into the representations of $\rho(\mathfrak{su}(2)) \oplus \mathfrak{f}$,

\begin{displaymath}
ad_{\mathfrak{g}}=\bigoplus_{n}V_n \otimes R_{n}.
\end{displaymath}
Here $V_n$ is the $n$-dimensional representation of $\mathfrak{su}(2)$, and $R_n$ is a (possibly reducible) representation of $\mathfrak{f}$. Based on this decomposition, the K-factor \cite{Gadde:2011uv, Lemos:2012ph, Gaiotto:2012uq, Gaiotto:2012xa} is the plethystic exponential \cite{Feng:2007ur}

\begin{displaymath}
K(a(O)) = PE \left[ \sum_{n} \frac{\tau^{n+1} \chi^{\mathfrak{f}}_{R_n}(a(O))}{1-\tau^2} \right].
\end{displaymath}
We have checked this proposal by identifying the rank-0 and rank-1 theories among our fixtures, and comparing their (already known) indices with the formula \eqref{betaalphaalpha}. These fixtures can be identified by computing the $(a,c)$ anomaly coefficients (equivalently the effective number of hyper and vector multiplets $(n_h,n_v)$) of the fixtures and the dimensions of Coulomb branch operators using formulas in \cite{Chacaltana:2012zy}. For rank-0 theories, i.e. free-field fixtures, $n_h$ determines the SCFT fully, while for rank-1 theories, the knowledge of anomaly coefficients $(a,c)$ along with the dimension of the single generator of the Coulomb branch chiral ring determines the SCFT uniquely \cite{Argyres:2015ffa,Argyres:2015gha,Argyres:2016xmc}.

As an example, consider the fixture labeled by $(0_\beta, [1^6]_\alpha, [6]_{\alpha\beta})$. It has one Coulomb branch operator of scaling dimension 4 and $(n_h,n_v)=(24,7)$ (or $(a,c)=(\frac{59}{24},\frac{19}{6})$). This fixture \emph{must} be the rank-1 ${(E_7)}_8$ Minahan-Nemeschansky theory \cite{Minahan:1996cj} whose index is already known \cite{Benvenuti:2010pq,Gadde:2011uv}. We have checked the unrefined index \eqref{betaalphaalpha} against the known index for the $(E_7)_8$ Minahan-Nemeschansky theory (up to order $\tau^{10}$), and found perfect agreement.

\subsection{$(\mathbb{1}, [\beta], [\beta])$ sector}\label{_sector_3}

The superconformal index of $(\mathbb{1},\beta,\beta^2)$ sector was studied in \cite{Chacaltana:2016shw}. There, this index was written as a sum over Dynkin labels of $\mathfrak{so}(8)$ representations that are invariant under the action of outer automorphism group $S_3$. It follows that the action of $C_{h_1, h_2, h_3}$ leaves the superconformal index invariant, and hence, the choice of internal twists and/or the representatives of $[\beta]$ conjugacy class does not affect the physics of the fixture.

\subsection{$(\mathbb{1}, [\alpha], [\alpha])$ sector}\label{_sector_4}

The superconformal index of a fixture in the $(\mathbb{1}, \alpha, \alpha)$ sector with trivial internal twists was studied in \cite{Lemos:2012ph}. The Schur limit of this index is

\begin{subequations}
\begin{equation}\label{1aaind}
\begin{split}
I^{(\mathbb{1},\alpha,\alpha,\mathbb{1},\mathbb{1},\mathbb{1})}_{Schur}& (\tau) = \frac{K(a(O_1), \tau)K(b(O_2), \tau) K(c(O_3), \tau )}{K([7,1],\tau)} \\
&\times  \sum_{\mathclap{(n_1, n_2, n_3)}} \frac{ \chi_{SO(8)}^{(n_1, n_2, n_3, n_3)} (a(O_1), \tau) \chi_{Sp(3)}^{(n_1, n_2, n_3)} (b(O_2), \tau)
 \chi_{Sp(3)}^{(n_1, n_2, n_3)}(c(O_3), \tau)
}
{ \chi_{SO(8)}^{(n_1, n_2, n_3, n_3)} ([7,1],\tau)}
\end{split}
\end{equation}
The Dynkin labels of $\mathfrak{so}(8)$ representations appearing in this formula are of the form $(n_1, n_2, n_3 , n_3)$, i.e. the index gets contribution only from the $\alpha$-invariant highest weights in the $\mathfrak{so}(8)$ weight space.

The indices in the two other sectors with trivial internal twists and $\gamma_2=\gamma_3=\alpha\beta^j$ are related to \eqref{1aaind} by the obvious $\mathfrak{so}(8)$ triality,

\begin{equation}\label{1ababind}
\begin{split}
I^{(\mathbb{1},\alpha\beta,\alpha\beta,\mathbb{1},\mathbb{1},\mathbb{1})}_{Schur}& (\tau) = \frac{K(a(O_1), \tau) K(b(O_2), \tau) K(c(O_3), \tau ) }{K([7,1],\tau)}\\
&\times \sum_{\mathclap{(n_1, n_2, n_3)}} \frac{ \chi_{SO(8)}^{(n_3, n_2, n_1, n_3)} (a(O_1), \tau) \chi_{Sp(3)}^{(n_1, n_2, n_3)} (b(O_2), \tau)
 \chi_{Sp(3)}^{(n_1, n_2, n_3)}(c(O_3), \tau)
}
{ \chi_{SO(8)}^{(n_3, n_2, n_1, n_3)} ([7,1],\tau)}
\end{split}
\end{equation}
and

\begin{equation}\label{1ab2ab2ind}
\begin{split}
I^{(\mathbb{1},\alpha\beta^2,\alpha\beta^2,\mathbb{1},\mathbb{1},\mathbb{1})}_{Schur}& (\tau) = \frac{K(a(O_1), \tau) K(b(O_2), \tau) K(c(O_3), \tau )}{K([7,1],\tau)}\\
&\times \sum_{\mathclap{(n_1, n_2, n_3)}} \frac{ \chi_{SO(8)}^{(n_3, n_2, n_3, n_1)} (a(O_1), \tau) \chi_{Sp(3)}^{(n_1, n_2, n_3)} (b(O_2), \tau)
 \chi_{Sp(3)}^{(n_1, n_2, n_3)}(c(O_3), \tau)
}
{ \chi_{SO(8)}^{(n_3, n_2, n_3, n_1)} ([7,1],\tau)}.
\end{split}
\end{equation}
\end{subequations}

\noindent We think of these sectors as being obtained from \eqref{1aaind} by acting with the conjugation $C_{h,h,h}$. $h=\alpha\beta^2$ or $\beta$ maps $(\mathbb{1},\alpha,\alpha,\mathbb{1},\mathbb{1},\mathbb{1})$ to $(\mathbb{1},\alpha\beta,\alpha\beta,\mathbb{1},\mathbb{1},\mathbb{1})$, i.e. takes \eqref{1aaind} to \eqref{1ababind}. Similarly, $h=\alpha\beta$ or $\beta^2$ takes \eqref{1aaind} to \eqref{1ab2ab2ind}.

$C_{\mathbb{1},h_2,h_3}$ does not act on the $SO(8)$ weights at the first puncture and acts trivially on the $Sp(3)$ weights at the other punctures. So it leaves the RHS of \eqref{1aaind}, \eqref{1ababind}, \eqref{1ab2ab2ind} invariant, while changing the twists. For instance, acting on \eqref{1aaind} with $C_{\mathbb{1},\mathbb{1},\beta}$, $C_{\mathbb{1},\mathbb{1},\alpha\beta^2}$, $C_{\mathbb{1},\alpha,\beta}$ and $C_{\mathbb{1},\alpha,\alpha\beta^2}$, respectively, we learn that
\begin{align*}
I^{(\mathbb{1},\alpha,\alpha\beta,\mathbb{1},\beta,\beta^2)}_{Schur} (\tau)&=
I^{(\mathbb{1},\alpha,\alpha\beta,\mathbb{1},\alpha\beta^2,\beta^2)}_{Schur} (\tau) =I^{(\mathbb{1},\alpha,\alpha\beta,\alpha,\alpha\beta^2,\beta^2)}_{Schur} (\tau)\\
&=
I^{(\mathbb{1},\alpha,\alpha\beta,\alpha,\beta,\alpha\beta^2)}_{Schur} (\tau) =
I^{(\mathbb{1},\alpha,\alpha,\mathbb{1},\mathbb{1},\mathbb{1})}_{Schur} (\tau).
\end{align*}

\noindent If, instead, we act on \eqref{1ababind} with $C_{\mathbb{1},      \beta^2,\mathbb{1}}$, $C_{\mathbb{1},\alpha\beta^2,\mathbb{1}}$, $C_{\mathbb{1},      \beta^2,\alpha\beta}$ or $C_{\mathbb{1},\alpha\beta^2,\alpha\beta}$, we find
\begin{align*}
I^{(\mathbb{1},\alpha,\alpha\beta,      \beta^2,        \beta, \mathbb{1})}_{Schur} (\tau)&=
I^{(\mathbb{1},\alpha,\alpha\beta,\alpha\beta^2,\alpha\beta^2, \mathbb{1})}_{Schur} (\tau) =
I^{(\mathbb{1},\alpha,\alpha\beta,      \beta^2,\alpha\beta^2,\alpha\beta)}_{Schur} (\tau)\\
&=
I^{(\mathbb{1},\alpha,\alpha\beta,\alpha\beta^2,        \beta,\alpha\beta)}_{Schur} (\tau) =
I^{(\mathbb{1},\alpha\beta,\alpha\beta,\mathbb{1},\mathbb{1},\mathbb{1})}_{Schur} (\tau).
\end{align*}
In fact, the index depends only on the isomorphism class of the $S_3$ bundle. Hence there is an easy invariant that determines which of the formulae \eqref{1aaind}, \eqref{1ababind} or \eqref{1ab2ab2ind} to use. For any fixture in this sector, consider a loop based at the untwisted puncture that encloses one of the two twisted punctures. It takes values $\gamma_4^{-1} \gamma_2 \gamma_4=\gamma_6 \gamma_3 \gamma_6^{-1}=\alpha \beta^{j}$ for some $j \in {0,1,2}$. $j=0$ corresponds to \eqref{1aaind}, $j=1$ corresponds to \eqref{1ababind} and $j=2$ corresponds to \eqref{1ab2ab2ind}. The value of $j$ is invariant under the action of $C_{\mathbb{1},h_2,h_3}$ for arbitrary $h_2$ and $h_3$, so we can simply use the value of $j$ to distinguish these formulae.

\subsection{$(\mathbb{1},\mathbb{1},\mathbb{1})$ sector}\label{_sector_5}

The superconformal index of this sector has been studied in \cite{Lemos:2012ph} in the special case where all the internal twists are trivial. In Schur limit, it reads

\begin{equation}\label{111ind}
\begin{split}
I^{(\mathbb{1},\mathbb{1},\mathbb{1},\mathbb{1},\mathbb{1},\mathbb{1})}_{Schur}& (\tau) =  \frac{K(a(O_1), \tau) K(b(O_2), \tau) K(c(O_3), \tau) }{K([7,1],\tau)} \\
&\times \sum_{\mathclap{(n_1, n_2, n_3,n_4)}} \frac{\chi_{SO(8)}^{(n_1, n_2, n_3, n_4)} (a(O_1), \tau) \chi_{SO(8)}^{(n_1, n_2, n_3, n_4)} (b(O_2), \tau)
 \chi_{SO(8)}^{(n_1, n_2, n_3, n_4)}(c(O_3), \tau)
}
{ \chi_{SO(8)}^{(n_1, n_2, n_3, n_4)} ([7,1],\tau)}.
\end{split}
\end{equation}
Here $(n_1, n_2, n_3, n_4)$ are the Dynkin labels of irreducible representations of $\mathfrak{so}(8)$ and $a(O_i)$ denote flavor fugacities of an $\mathfrak{so}(8)$ nilpotent orbit $O_i$. These are to be expressed in terms of $\tau$ and the fugacities for $\mathfrak{f}_i$, the manifest flavor symmetry of the puncture.

The index in the presence of internal twists $(\gamma_4, \gamma_5, \gamma_6) = (h_2 h_1^{-1}, h_3 h_2^{-1}, h_1 h_3^{-1})$ is related to this formula by the action of conjugation operator $C_{h_1,h_2,h_3}$.
\begin{equation}\label{111indconjugated}
\begin{split}
&I^{(\mathbb{1},\mathbb{1},\mathbb{1},h_2 h_1^{-1}, h_3 h_2^{-1}, h_1 h_3^{-1})}_{Schur} (\tau) = \frac{K(a(O_1), \tau)K(b(O_2), \tau)K(c(O_3), \tau)}{K([7,1],\tau)} \\
&\times \sum_{\mathclap{(n_1, n_2, n_3,n_4)}} \frac{ \chi_{SO(8)}^{h_1 \cdot (n_1, n_2, n_3, n_4)} (a(O_1), \tau) \chi_{SO(8)}^{h_2 \cdot (n_1, n_2, n_3, n_4)} (b(O_2), \tau)
 \chi_{SO(8)}^{h_3 \cdot(n_1, n_2, n_3, n_4)}(c(O_3), \tau)
}{\chi_{SO(8)}^{(n_1, n_2, n_3, n_4)} ([7,1],\tau)}
\end{split}
\end{equation}

\noindent Here $h_i \cdot (n_1, n_2, n_3, n_4)$ denotes the action of $h_i$ on the highest weight. The choice of conjugation operator is not unique; different choices are related by sending $h_i \to h_i g$ for $i =1,2,3$. This change amounts to an action on the index by a global conjugation operator $C_{g,g,g}$, which leaves the index \eqref{111ind} invariant.

The action of an outer automorphism on weights in expression \eqref{111indconjugated} may be exchanged with an action on the fugacities at which the characters are evaluated. In particular, $\chi_{SO(8)}^{h_i \cdot (n_1, n_2, n_3, n_4)} (a(O_i), \tau)= \chi_{SO(8)}^{ (n_1, n_2, n_3, n_4)} (h_i^{-1} \cdot (a(O_i), \tau))$. Since fugacities are determined by the embedding $\rho: \mathfrak{su}(2) \rightarrow \mathfrak{so}(8)$ that determines the nilpotent orbit $O_i$, this action just amounts to composing $\rho$ with $h_i^{-1}$ to get another embedding $h_i^{-1}\circ \rho$.

Depending on the choice of $O$, this action might preserve the $\mathfrak{su}(2)$ embedding and just act on the flavour fugacities. Or it might change the $\mathfrak{su}(2)$ embedding --- i.e., change the nilpotent orbit $O$. $O=[1^8]$ and $O=[2^2,1^4]$ are examples of the former. In the case of $[1^8]$, the outer automorphism acts on the fugacities in a way that permutes the three 8-dimensional representations, $8_{v,s,c}$; in the case of $[2^2,1^4]$, it permutes the fugacities associated to $\mathfrak{f}=\mathfrak{su}(2) \oplus \mathfrak{su}(2) \oplus  \mathfrak{su}(2) $. The triple of orbits, $O={\color{darkgreen}[3,1^5]},{\color{red}[2^4]},{\color{blue}[2^4]}$, are an example of the latter. In each of these embeddings, \emph{one} of the 8-dimensional representations decomposes as $(3,1)+(1,5)$ of $\mathfrak{su}(2) \oplus \mathfrak{sp}(2)$ and the other two decompose as $(2,4)$. The outer automorphism permutes the three 8-dimensional representations, and hence permutes the three punctures ${\color{darkgreen}[3,1^5]},{\color{red}[2^4]},{\color{blue}[2^4]}$.

By a suitable choice of conjugation, $C_{\mathbb{1},h_2,h_3}$, we can always remove the internal twists and thereby relate the fixture to one with trivial internal twists. For example, consider\footnote{\label{ft:colorpunctures}Throughout this paper, we specify external twists by filling in punctures with different colors. When the color is white, the puncture is untwisted. When it is green, red or blue, the twist is $\alpha$, $\alpha \beta$ or $\alpha \beta^2$ respectively, and when it is light grey or dark grey, it denotes a $\beta$ or $\beta^2$ twist.}

\begin{displaymath}
 \includegraphics[width=88pt]{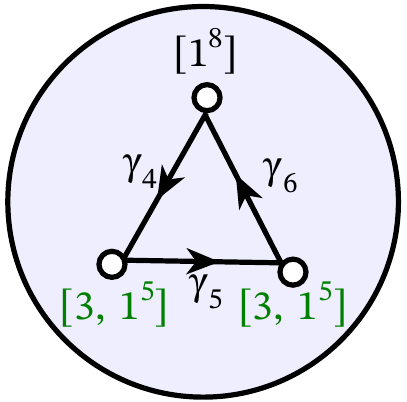}
\end{displaymath}
Acting with $C_{\mathbb{1},\gamma_4^{-1},\gamma_6}$ eliminates the internal twists, at the cost of some action on the punctures. If, for instance, $\gamma_4 = \alpha$ and $\gamma_6=\beta$, this fixture is equivalent to

\begin{displaymath}
 \includegraphics[width=88pt]{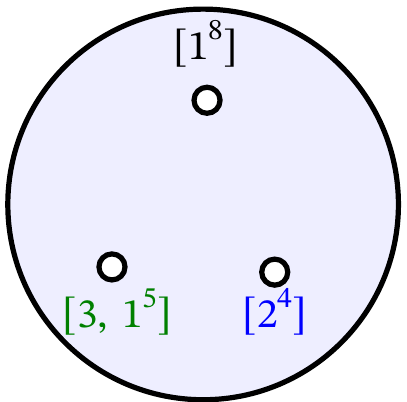}
\end{displaymath}
with trivial internal twists.

\section{Once-punctured torus}\label{oncepunctured_torus}

As we saw in the $(\mathbb{1},\mathbb{1},\mathbb{1})$ and in the $(\mathbb{1},[\alpha],[\alpha])$ sector, different choices of internal twists can yield non-isomorphic physics already at the level of the 3-punctured spheres. But even when the SCFTs associated to fixtures with different choices of internal twists \emph{are} isomorphic, they can lead to distinct physics once we start gauging the global symmetries. The simplest examples arise when we sew together two punctures on a 3-punctured sphere to obtain a once-punctured torus.

Consider, first, the case where the remaining puncture is in the untwisted sector. The twists around the $a$ and $b$ cycles of the torus must commute. There are 18 possibilities, which break up into 5 orbits under the modular group.

\begin{equation}
\begin{gathered}
(\mathbb{1},\mathbb{1})\\
\{ (\mathbb{1},\alpha), (\alpha,\mathbb{1}), (\alpha,\alpha)\}\\
\{ (\mathbb{1},\alpha\beta), (\alpha\beta,\mathbb{1}), (\alpha\beta,\alpha\beta)\}\\
\{ (\mathbb{1},\alpha\beta^2), (\alpha\beta^2,\mathbb{1}), (\alpha\beta^2,\alpha\beta^2)\}\\
\{ (\mathbb{1},\beta), (\mathbb{1},\beta^2), (\beta,\mathbb{1}), (\beta,\beta), (\beta,\beta^2), (\beta^2,\mathbb{1}),(\beta^2,\beta),(\beta^2,\beta^2)\}
\end{gathered}
\label{torusorbits}\end{equation}
For each orbit, there is (at least) one element where the twist around the $a$-cycle is $\mathbb{1}$, so we can think of the theory as being obtained from a 3-punctured sphere with two untwisted full punctures and one other untwisted puncture, by gauging the diagonal $Spin(8)$ associated to the full punctures. If the full punctures that we sew together are 1 and 2, then the twist around the $b$-cycle on the sewn surface is the internal twist, $\gamma_4$ which connects them. $\gamma_4$ can be any element of $S_3$, so we can arrive at any of the 5 orbits listed above.

Each orbit in \eqref{torusorbits} is a fibration over $\overline{\mathcal{M}}_{1,1}$, the compactification of the fundamental domain for $PSL(2,\mathbb{Z})$. The first orbit is just $\overline{\mathcal{M}}_{1,1}$ itself. The next 3 are copies of (the compactification of) the fundamental domain of $\tilde\Gamma_0(2)$ and the last is (the compactification of) the fundamental domain\footnote{Up to conjugation, the congruence subgroups of $SL(2,\mathbb{Z})$ are

\begin{displaymath}
\begin{aligned}
\Gamma(n)=\left\{\begin{pmatrix}1&0\\0&1\end{pmatrix}\; \mod n
\right\}\subset SL(2,\mathbb{Z})\\
\Gamma_1(n)=\left\{\begin{pmatrix}1&b\\0&1\end{pmatrix}\; \mod n
\right\}\subset SL(2,\mathbb{Z})\\
\Gamma_0(n)=\left\{\begin{pmatrix}a&b\\0&d\end{pmatrix}\; \mod n
\right\}\subset SL(2,\mathbb{Z})\\
\end{aligned}
\end{displaymath}
where, obviously, we have $\Gamma(n)\subset\Gamma_1(n)\subset\Gamma_0(n)\subset SL(2,\mathbb{Z})$ and $\Gamma_1(2)\simeq\Gamma_0(2)$. The center $\mathbb{Z}_2\subset SL(2,\mathbb{Z})$ acts trivially on the Upper Half-Plane; the quotient $PSL(2,\mathbb{Z})=SL(2,\mathbb{Z})/\mathbb{Z}_2$ acts effectively. $\Gamma(2)$ and $\Gamma_0(2)$ contain the center; let us denote the quotients by $\tilde\Gamma(2)=\Gamma(2)/\mathbb{Z}_2$ and $\tilde\Gamma_0(2)=\Gamma_0(2)/\mathbb{Z}_2$. Several of the moduli spaces that crop up in our analysis are quotients of the UHP by one of these congruence subgroups: $\mathcal{M}_{0,4}= \text{UHP}/\tilde\Gamma(2)$, $\mathcal{M}_{1,1}=\text{UHP}/PSL(2,\mathbb{Z})$  and  $\text{UHP}/\Gamma_1(p)$ (for $p>2$) is the moduli space of pairs $(T,\gamma)$, consisting of a torus, $T$, and a nonzero element $\gamma\in H^1(T,\mathbb{Z}/p)$.}  of $\Gamma_1(3)$. In both cases, the compactification is ramified over the boundary point of $\overline{\mathcal{M}}_{1,1}$. In the $\tilde\Gamma_0(2)$ case, two of the three sheets of the covering come together over the boundary, so there are two distinct weak-coupling limits, depending on whether the twist around the shrinking cycle is $\mathbb{1}$ or in $[\alpha]$ conjugacy class. In the $\Gamma_1(3)$ case, the four sheets come together to form a 2:1 cover of the boundary; the distinct weak coupling limits correspond, respectively, to the twist around the shrinking cycle being in $[\beta]$ conjugacy class or $\mathbb{1}$.

As our first example, let puncture 3 be $[5,3]$, so that the 3-punctured sphere, before gauging, is the $(E_8)_{12}$ Minahan-Nemeschansky SCFT \cite{Minahan:1996cj}.

\begin{equation}
\begin{matrix}
 \includegraphics[width=130pt]{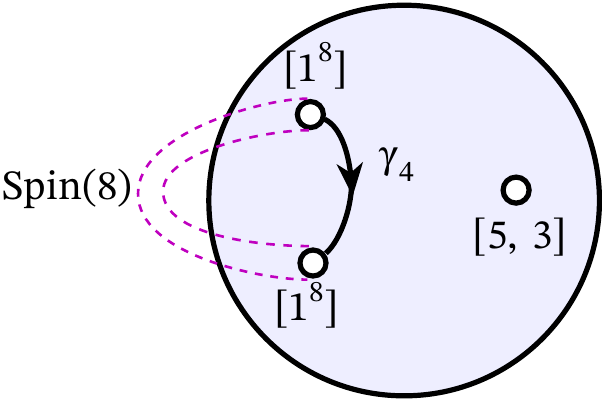}
 \end{matrix}
\label{untwistedE8MN}\end{equation}
Different choices of $\gamma_4$ correspond to different embeddings of the $Spin(8)$ gauge group in the manifest $Spin(8)\times Spin(8)\subset E_8$. Under the ``standard'' embedding of $Spin(8)\times Spin(8)$, the adjoint of $E_8$ decomposes as $(28,1) \oplus (1,28) \oplus (8_v, 8_v) \oplus (8_c, 8_c)  \oplus (8_s, 8_s)$. If $\gamma_4=\mathbb{1}$, then gauge group embeds diagonally in the above, and the $248$ decomposes as

\begin{displaymath}
\begin{aligned}
  248&= 28 \oplus 28 \oplus (8_v\otimes 8_v)\oplus (8_s\otimes 8_s)\oplus (8_c\otimes 8_c)\\
&=28\oplus 28\oplus(28\oplus35_v\oplus1)\oplus(28\oplus35_s\oplus1)\oplus(28\oplus35_c\oplus1).
\end{aligned}
\end{displaymath}
The three singlet $\hat{B}_1$ operators survive the gauging, and so the flavour symmetry is $U(1)^3$.

We can obtain other values of $\gamma_4$ by acting with the conjugation $C_{h_1,\mathbb{1},\mathbb{1}}$ as in \eqref{conjugation}. This sets $\gamma_4=h_1^{-1}$. The conjugation acts as an outer-automorphism of the $Spin(8)$ associated to the first puncture, and hence changes the decomposition of the $248$ (relative to what we had before, with $\gamma_4=\mathbb{1}$). If $h_1=\alpha$, the $248$ decomposes as $(28,1) \oplus (1,28) \oplus (8_v, 8_v) \oplus (8_s, 8_c)  \oplus (8_c, 8_s)$ and so under the $Spin(8)$ gauging, we have

\begin{displaymath}
248=28\oplus28\oplus(28\oplus35_v\oplus1)\oplus(8_v\oplus56_v)\oplus(8_v\oplus56_v)
\end{displaymath}
Only one $\hat{B}_1$ operator survives the gauging, and the flavour symmetry is $U(1)$.

The S-dual descriptions involve replacing the $[1^8]$ punctures from the untwisted sector with the $[1^6]$ punctures from the $\alpha$-twisted sector.

\begin{displaymath}
 \includegraphics[width=126pt]{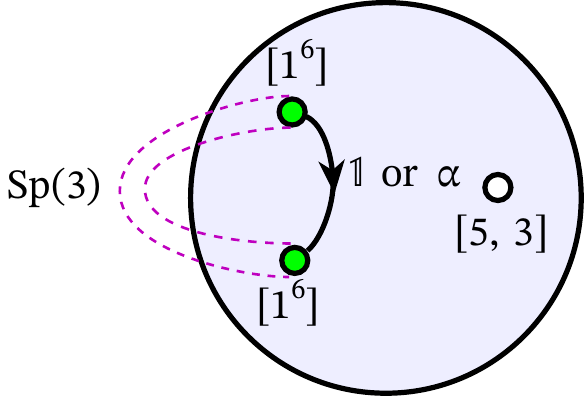}
\end{displaymath}
The conformal manifold (the fundamental domain for $\tilde\Gamma_0(2)$) is a 3-sheeted cover of $\overline{\mathcal{M}}_{1,1}$. Over the boundary of $\overline{\mathcal{M}}_{1,1}$, two of the sheets (the ones corresponding to the weakly-coupled $Sp(3)$ description) come together and there are only 2 distinct weakly-coupled descriptions. One is the $Spin(8)$-gauging of the ${(E_8)}_{12}$ Minahan-Nemeschansky theory and the other is the $Sp(3)$-gauging of the rank-2 ${Sp(6)}_8$ SCFT.

The same analysis holds for $\gamma_4=\alpha\beta$ and $\alpha\beta^2$: we get a family of SCFTs with a $U(1)$ flavour symmetry, whose conformal manifold is (the compactification of) the fundamental domain for $\tilde\Gamma_0(2)$, fibered over $\mathcal{M}_{1,1}$, and whose two weakly-coupled descriptions are a $Spin(8)$-gauging of the ${(E_8)}_{12}$ Minahan-Nemeschansky theory and an $Sp(3)$-gauging of the ${Sp(6)}_8$ SCFT.

Finally, if $\gamma_4=\beta$, we get a cyclic permutation of the three 8-dimensional representations of the $Spin(8)$ associated to the first puncture and upon gauging,

\begin{displaymath}
248=28\oplus28\oplus(8_v\oplus56_v)\oplus(8_s\oplus56_s)\oplus(8_c\oplus56_c).
\end{displaymath}
No $\hat{B}_1$ operators survive the gauging and the flavour symmetry is trivial.

In the same orbit, we can replace the $[1^8]$ untwisted punctures by the full punctures from the $\beta$- or $\beta^2$-twisted sector. The corresponding fixture is the rank-3 ${(G_2)}_8^2$ SCFT and the once-punctured torus theory is its diagonal $G_2$-gauging.

\begin{displaymath}
 \includegraphics[width=115pt]{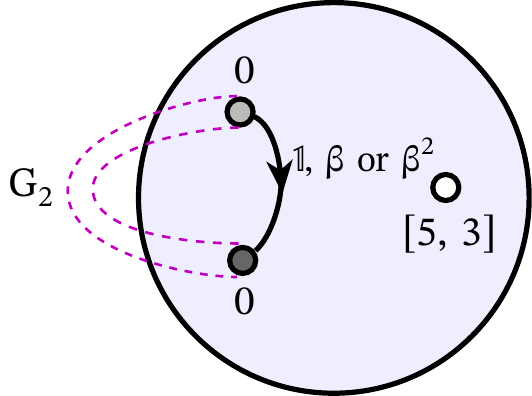}
\end{displaymath}
The conformal manifold is (the compactification of) the fundamental domain of $\Gamma_1(3)$, a $4$-sheeted cover of $\overline{\mathcal{M}}_{1,1}$. Over the boundary, there are two weakly-coupled descriptions: a $Spin(8)$ gauging of the ${(E_8)}_{12}$ SCFT (if the twist around the shrinking cycle is $\mathbb{1}$) and the $G_2$ gauging of the ${(G_2)}_8^2$ SCFT (if the twist is in the conjugacy class $[\beta]$).

Thus, of the 5 orbits in \eqref{torusorbits}:

\begin{itemize}%
\item The first has a ${U(1)}^3$ flavour symmetry and has a unique weakly-coupled description as a $Spin(8)$-gauging of the ${(E_8)}_{12}$ SCFT.
\item The next three are isomorphic families of SCFTs, with a $U(1)$ flavour symmetry and two weakly-coupled descriptions, as a (different) $Spin(8)$-gauging of the ${(E_8)}_{12}$ SCFT or as an $Sp(3)$-gauging of the ${Sp(6)}_8$ SCFT.
\item The last is a family of SCFTs with trivial flavour symmetry, with two weakly-coupled descriptions, as a (third) $Spin(8)$-gauging of the ${(E_8)}_{12}$ SCFT or as a $G_2$-gauging of the ${(G_2)}_8^2$ SCFT.

\end{itemize}
Starting with the fixture \eqref{untwistedE8MN}, which of these three cases is obtained depends on the choice of (conjugacy class of) $\gamma_4$. The first corresponds to $\gamma_4=\mathbb{1}$, the second to $[\gamma_4]=[\alpha]$ and the third to $[\gamma_4]=[\beta]$.

For our second example, consider replacing the $[5,3]$ puncture by the puncture $\color{darkgreen}[5,1^3]$ that lies in a triplet of $S_3$ (along with $\color{red}[4^2]$ and $\color{blue}[4^2]$). Before gauging, the fixture is a $Spin(16)_{12} \times SU(2)_8$ SCFT.

\begin{equation}\begin{matrix}
 \includegraphics[width=130pt]{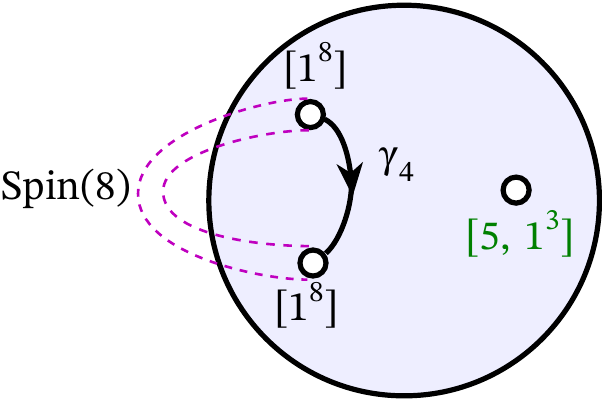}
 \end{matrix}
\label{Spin16xSU2}\end{equation}
As before, this figure labels entries $(\mathbb{1}, \gamma_4)$ in \eqref{torusorbits}. If $\gamma_4 = \mathbb{1}$, the gauge group embeds diagonally, and the adjoint of $Spin(16)$ decomposes as

\begin{equation}
\begin{aligned}
  120&= 28 \oplus 28 \oplus (8_v\otimes 8_v)\\
&=28\oplus 28\oplus28\oplus35_v\oplus1
\end{aligned}
\label{decomposition}\end{equation}
The singlet $\hat{B}_1$ operator survives the gauging, and the flavor symmetry of the gauge theory is $SU(2) \times U(1)$. There is a unique weakly-coupled description, and the conformal manifold is the fundamental domain of $PSL(2,\mathbb{Z})$.

When $\gamma_4 = \alpha$, the decomposition is the same as in \eqref{decomposition}, and therefore, the flavor symmetry is the same. However, there is now an S-dual frame.

\begin{displaymath}
 \includegraphics[width=126pt]{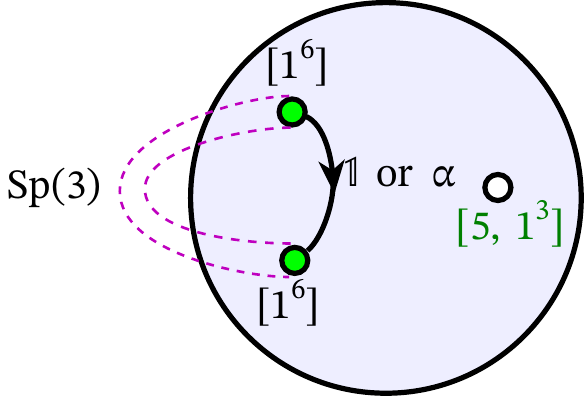}
\end{displaymath}
The conformal manifold is the fundamental domain of $\tilde\Gamma_0 (2)$, with two weakly-coupled limits. These are a $Spin(8)$-gauging of the $Spin(16)_{12} \times SU(2)_8$ SCFT and an $Sp(3)$-gauging of the $Sp(6)_8 \times SU(2)_8$ SCFT.

When $\gamma_4 = \alpha \beta$ or $\alpha \beta^2$, the decomposition \eqref{decomposition} changes to $120= 28 \oplus 28 \oplus 8_{s/c} \otimes 8_v$. Therefore, the flavor symmetry of the gauge theory changes to $SU(2)$ instead of $SU(2) \times U(1)$. If $\gamma_4 = \alpha \beta$, the S-dual description is

\begin{displaymath}
 \includegraphics[width=126pt]{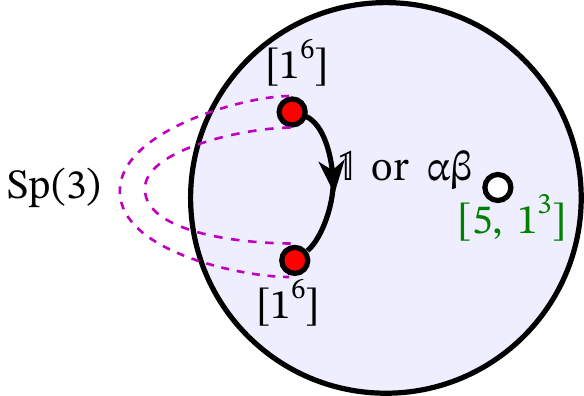}
\end{displaymath}
If $\gamma_4 = \alpha \beta^2$, the red puncture is replaced by the blue puncture, and the twist line can take values $\mathbb{1}$ or $\alpha \beta^2$. The conformal manifold in either case is the fundamental domain of $\tilde\Gamma_0(2)$ with two weakly-coupled limits. These are a $Spin(8)$-gauging of the $Spin(16)_{12}\times SU(2)_8$ SCFT and an $Sp(3)$-gauging of the $Sp(3)_8^2 \times SU(2)_8$ SCFT.

Finally, if $\gamma_4 = \beta$ or $\beta^2$, the flavor symmetry remains $SU(2)$, but the conformal manifold of the gauge theory changes. In particular, the S-dual description is

\begin{displaymath}
 \includegraphics[width=115pt]{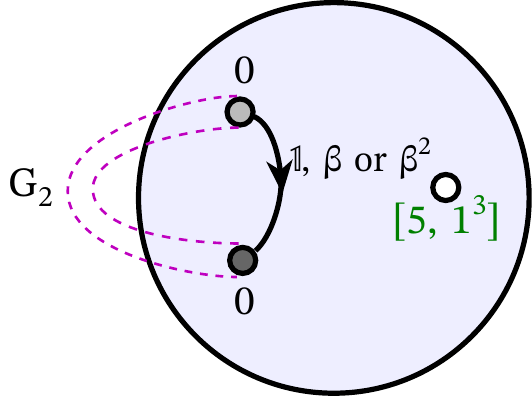}
\end{displaymath}
The conformal manifold is the fundamental domain of $\Gamma_1(3)$. It has two weakly-coupled descriptions: a $Spin(8)$-gauging of the $Spin(16)_{12}\times SU(2)_8$ SCFT and a $G_2$-gauging of the $(G_2)_8^2 \times SU(2)_8$ SCFT.

By contrast, the physics of the 3-punctured spheres in the $([\alpha],[\alpha],[\beta])$ and $([\beta],[\beta],[\beta])$ sectors is supposed to be independent of the internal twists. Consider the once-punctured torus, with the puncture in the $\beta$-twisted sector. There are $18$ choices for the twists around the $a$- and $b$-cycles of the torus; any pair of non-commuting elements of $S_3$ will do. These 18 choices form a single orbit under the modular group. Up to an overall conjugation by an element of $S_3$, this yields a conformal manifold which is the 3-sheeted cover of $\mathcal{M}_{1,1}$, $\text{UHP}/\tilde\Gamma_0(2)$. Over the boundary in $\overline{\mathcal{M}}_{1,1}$, two sheets come together and there are two physically-distinct weak-coupling limits, depending on whether the twist around the shrinking cycle is in the $[\alpha]$ or the $[\beta]$ conjugacy class.

Take, for instance, the $A_1$ puncture from the $\beta$-twisted sector. If the twist around the shrinking cycle of the torus is in the conjugacy class $[\alpha]$, then we get an $Sp(3)$ gauging of the ${Sp(3)}_8^2\times {SU(2)}_{14}$ SCFT; if the twist is in the $[\beta]$ conjugacy class, we get a $G_2$ gauging of the ${(G_2)}_8^2\times {SU(2)}_{14}$ SCFT:

\begin{displaymath}
\begin{matrix}
 \includegraphics[width=126pt]{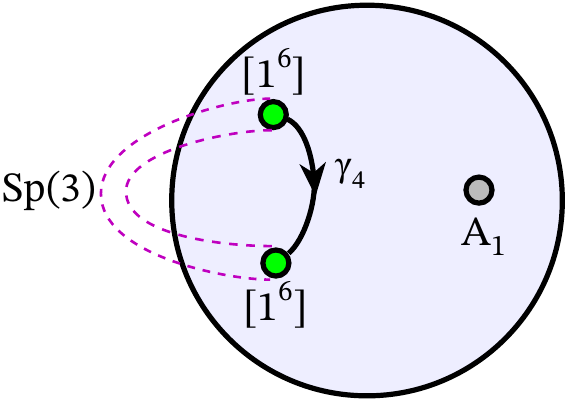}
\end{matrix}\quad\text{or}\quad
\begin{matrix}
 \includegraphics[width=115pt]{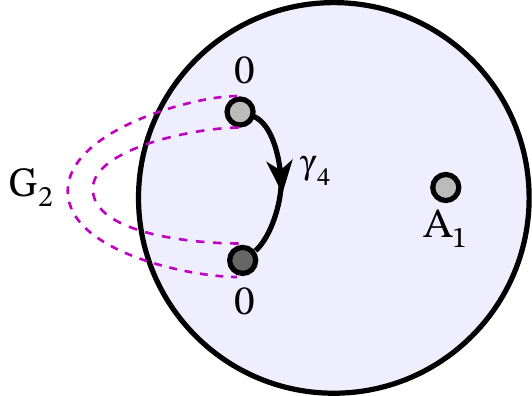}
\end{matrix}
\end{displaymath}

\section{Classification of fixtures}\label{classification_of_fixtures}

Here, we use the various formulas for the superconformal index from \S\ref{superconformal_index_for_nonabelian_twisted__theories} to classify fixtures in the corresponding sectors.\footnote{Before we present the classification, we should make a remark about fixtures with irregular punctures here. Recall that in \S\ref{superconformal_index_for_nonabelian_twisted__theories}, we argued that we can relate some of the new theories with internal twists to theories without internal twists which have been classified before. This argument uses various conjugations operators and the action of outer automorphisms on nilpotent orbits. We assume this also holds for fixtures with irregular punctures, so that we can relate theories involving internal twists to those without them. Since we do not have a formula for the index of an irregular fixture, we cannot directly prove this, but the consistency of various S-dualities requires it to be true.}  Here is a brief summary of the new theories in each sector.

\begin{enumerate}%
\item In the $([\beta],[\beta],[\beta])$ sector, there is one irregular fixture that was not previously discussed in \cite{Chacaltana:2016shw}, and we will list this fixture \S\ref{classification_of_fixtures_in__sector}.
\item The $([\beta],[\alpha],[\alpha])$ sector has not been studied previously. We will present a complete classification of fixtures in this sector in \S\ref{classification_of_fixtures_in__sector_2}.
\item The $(\mathbb{1},[\beta],[\beta])$ sector has no new theories. This sector was studied in \cite{Chacaltana:2016shw}.
\item In the $(\mathbb{1},\mathbb{1},\mathbb{1})$ sector, one might expect new theories due to the non-trivial action of conjugation on the untwisted punctures, but as we argued in \S\ref{_sector_5}, any fixture in this sector with internal twists is isomorphic to a fixture without internal twists possibly with some punctures replaced. This implies that any theory one comes across is already present in the tables given in \cite{Chacaltana:2011ze}.
\item Finally, in the $(\mathbb{1},[\alpha],[\alpha])$ sector, we have argued in \S\ref{_sector_4} that the twist around a loop based at the untwisted puncture that contains a twisted puncture, determines the physics of the fixture. There are three possible values for this twist, and any theory one finds is related to a theory in \cite{Chacaltana:2013oka} (where the twist around the loop is $\alpha$) via triality.

\end{enumerate}

\subsection{Classification of fixtures in $([\beta],[\beta],[\beta])$ sector} \label{classification_of_fixtures_in__sector}

As we remarked above, $(\beta, \beta, \beta)$ sector was studied in \cite{Chacaltana:2016shw} where fixtures of this sector were also classified. We have found one more irregular fixture which corresponds to a gauge theory, and we present it below.

\begin{longtable}{|c|c|c|c|c|c|}
\hline
\#&Fixture&$(d_2,d_3,d_4,d_6)$&$G$&Num. of Hypers&Representation\\
\endhead
\hline 
1&$\begin{matrix}G_2\\ G_2\end{matrix}\quad (\tilde{A}_1,SU(2)_3)$&$(1, 0, 0, 0)$&$SU(2)$&$3$&$\tfrac{1}{2}(3; 2)$\\
\hline
\captionsetup{width=15cm}
\caption{The only fixture of the $([\beta],[\beta],[\beta])$ sector that was not listed in \cite{Chacaltana:2016shw}.}
\end{longtable} \label{irregfreefixtable}

\noindent Here $G$ is the gauge group and $d_i$ are the graded Coulomb branch dimensions for this gauge theory fixture.

\subsection{Classification of fixtures in $([\beta], [\alpha] , [\alpha])$ sector}\label{classification_of_fixtures_in__sector_2}

\subsubsection{Irregular fixtures}
\label{irregular_fixtures}

The $([\beta], [\alpha] , [\alpha])$ sector has eight irregular fixtures that correspond to free field theories, and three irregular fixtures that correspond to gauge theories.

\begin{longtable}{|c|c|c|c|}
\captionsetup{width=15cm}
\caption{Irregular fixtures of the $([\beta],[\alpha],[\alpha])$ sector corresponding to free SCFTs.}\\
\hline
\#&Fixture&Num. of hypers&Representation\\
\endfirsthead
\hline
\#&Fixture&Num. of hypers&Representation\\
\endhead
\hline
1&$\begin{matrix}[6]\\ [6]\end{matrix}\quad (G_2,\emptyset)$&$0$&empty\\
\hline
2&$\begin{matrix}[6]\\ [4, 2]\end{matrix}\quad (G_2(a_1),\emptyset)$&$0$&empty\\
\hline
3&$\begin{matrix}[6]\\ [4, 1^2]\end{matrix}\quad (\tilde{A}_1,\emptyset)$&$0$&empty\\
\hline
4&$\begin{matrix}[6]\\ [3^2]\end{matrix}\quad (\tilde{A}_1, {SU(2)}_3)$&$3$&$\tfrac{1}{2}(3, 2)$\\
\hline
5&$\begin{matrix}[6]\\ [2^3]\end{matrix}\quad (0, {SU(2)}_0)$&$0$&empty\\
\hline
6&$\begin{matrix}[6]\\ [2^2, 1^2]\end{matrix}\quad (0, {SU(3)}_4)$&$6$&$(2, 3)$\\
\hline
7&$\begin{matrix}[4, 1^2]\\ G_2\end{matrix}\quad ([1^6],{Sp(2)}_{4}\times {SU(2)}_0)$&$5$&$(5,1)$\\
\hline
8&$\begin{matrix}[6]\\ \tilde{A}_1\end{matrix}\quad ([1^6],{Sp(2)}_{4}\times {SU(2)}_0)$&$5$&$(5,1)$\\
\hline
\end{longtable} \label{irregfreefixtable}

{
\setlength\LTleft{-.125in}

\begin{longtable}{|c|c|c|c|c|c|}
\captionsetup{width=15cm}
\caption{Irregular fixtures of the $([\beta],[\alpha],[\alpha])$ sector corresponding to gauge theories.}\\
\hline
\#&Fixture&$(d_2,d_3,d_4,d_6)$&$G$&Num.  of Hypers&Representation\\
\endfirsthead
\hline
\#&Fixture&$(d_2,d_3,d_4,d_6)$&$G$&Num.  of Hypers&Representation\\
\endhead
\hline 
1&$\begin{matrix}[6]\\ G_2\end{matrix}\quad ([4,1^2],{SU(2)}_{3})$&$(1, 0, 0, 0)$&$SU(2)$&$3$&$\tfrac{1}{2}(3; 2)$\\ \hline
2&$\begin{matrix}[6]\\ G_2(a_1)\end{matrix}\quad ([2,1^4],{Sp(2)}_{5})$&$(2, 0, 0, 0)$&${SU(2)}^2$&$10$&$\tfrac{1}{2}(2, 2; 4)+\tfrac{1}{2}(1, 1; 4)$\\ \hline
3&$\begin{matrix}[4, 2]\\ G_2\end{matrix}\quad ([2,1^4],{Sp(2)}_{5})$&$(2, 0, 0, 0)$&${SU(2)}^2$&$10$&$\tfrac{1}{2}(2, 2; 4)+\tfrac{1}{2}(1, 1; 4)$\\
\hline
\end{longtable}
}

\subsubsection{Free field fixture}
\label{free_field_fixture}

There is only 1 regular fixture in $([\beta], [\alpha], [\alpha])$ sector which corresponds to a free field theory.

\begin{longtable}{|c|c|c|}
\hline
\#&Fixture&Representation\\
\endhead
\hline 
1&$\begin{matrix} [6] \\ [2,1^4]\end{matrix}\quad 0$&$\tfrac{1}{2}(4,7)$\\
\hline
\captionsetup{width=15cm}
\caption{Free field fixture of $([\beta],[\alpha],[\alpha])$ sector.}
\end{longtable}

\subsubsection{Interacting fixtures} \label{interacting_fixtures}

The $([\beta],[\alpha],[\alpha])$ sector has two interacting fixtures whose global symmetry is enhanced from their manifest symmetry.

\begin{longtable}{|c|c|c|c|c|}
\hline
\#&Fixture&$(d_3,d_4,d_6)$&$(n_h, n_v)$&$G_{\text{global}}$\\
\endhead
\captionsetup{width=15cm}
\caption{Interacting fixtures of $([\beta],[\alpha],[\alpha])$ sector with enhanced global symmetry.}\\
\endfoot
\hline
1&$\begin{matrix} [1^6] \\ [6]\end{matrix}\quad 0$&$(0, 1, 0)$&$(24, 7)$&$(E_7)_8$\\
\hline
2&$\begin{matrix} [4,1^2] \\ [4,1^2]\end{matrix}\quad 0$&$(0, 2, 0)$&$(26, 14)$&$Spin(7)_8 \times SU(2)_5^2$\\
\hline
\end{longtable}

\noindent Here $G_{\text{global}}$ is the global symmetry of an SCFT and $(n_h, n_v)$ denote the effective number of hyper and vector multiplets. This sector also has 28 interacting fixtures whose global symmetry is unenhanced. 

\begin{longtable}{|c|c|c|c|c|}
\captionsetup{width=16cm}
\caption{Interacting fixtures of $([\beta],[\alpha],[\alpha])$ sector with unenhanced global symmetry.}\\
\hline
\#&Fixture&$(d_3,d_4,d_6)$&$(n_h, n_v)$&$G_{\text{global}}$\\
\endfirsthead
\hline
\#&Fixture&$(d_3,d_4,d_6)$&$(n_h, n_v)$&$G_{\text{global}}$\\
\hline
\endhead
\endfoot
\endlastfoot
\hline
1&$\begin{matrix} [1^6] \\ [1^6]\end{matrix}\quad 0$&$(0, 6, 4)$&$(112, 86)$&$(G_2)_8^{} \times Sp(3)_8^2$\\ \hline
2&$\begin{matrix} [1^6] \\ [2,1^4]\end{matrix}\quad 0$&$(0, 5, 4)$&$(102, 79)$&$(G_2)_8^{} \times Sp(2)_7^{} \times Sp(3)_8^{}$\\  \hline
3&$\begin{matrix} [1^6] \\ [2^2,1^2]\end{matrix}\quad 0$&$(1, 5, 3)$&$(94, 73)$&$(G_2)_8^{} \times Sp(3)_8^{} \times SU(2)_6^{} \times U(1)^{}$\\  \hline
4&$\begin{matrix} [1^6] \\ [2^3]\end{matrix}\quad 0$&$(0, 5, 3)$&$(88, 68)$&$(G_2)_8^{} \times Sp(3)_8^{} \times SU(2)_{24}^{}$\\  \hline
5&$\begin{matrix} [1^6] \\ [4,1^2]\end{matrix}\quad 0$&$(0, 4, 2)$&$(69, 50)$&$(G_2)_8^{} \times Sp(3)_8^{} \times SU(2)_5^{}$\\  \hline
6&$\begin{matrix} [2,1^4] \\ [2,1^4]\end{matrix}\quad 0$&$(0, 4, 4)$&$(92, 72)$&$(G_2)_8^{} \times Sp(2)_7^2$\\  \hline
7&$\begin{matrix} [2,1^4] \\ [2^2,1^2]\end{matrix}\quad 0$&$(1, 4, 3)$&$(84, 66)$&$(G_2)_8^{} \times Sp(2)_7^{} \times SU(2)_6^{} \times U(1)^{}$\\  \hline
8&$\begin{matrix} [2,1^4] \\ [2^3]\end{matrix}\quad 0$&$(0, 4, 3)$&$(78, 61)$&$(G_2)_8^{} \times Sp(2)_7^{} \times SU(2)_{24}$\\  \hline
9&$\begin{matrix} [2,1^4] \\ [4,1^2]\end{matrix}\quad 0$&$(0, 3, 2)$&$(59, 43)$&$(G_2)_8^{} \times Sp(2)_7^{} \times SU(2)_5^{}$\\  \hline
10&$\begin{matrix} [2^2,1^2] \\ [2^2,1^2]\end{matrix}\quad 0$&$(2, 4, 2)$&$(76, 60)$&$(G_2)_8^{} \times SU(2)_6^2 \times U(1)^2$\\  \hline
11&$\begin{matrix} [2^2,1^2] \\ [2^3]\end{matrix}\quad 0$&$(1, 4, 2)$&$(70, 55)$&$(G_2)_8^{} \times SU(2)_{24} \times SU(2)_6^{} \times U(1)^{}$\\  \hline
12&$\begin{matrix} [2^2,1^2] \\ [4,1^2]\end{matrix}\quad 0$&$(1, 3, 1)$&$(51, 37)$&$(G_2)_8^{} \times SU(2)_5^{} \times SU(2)_6^{} \times U(1)^{}$\\  \hline
13&$\begin{matrix} [2^3] \\ [2^3]\end{matrix}\quad 0$&$(0, 4, 2)$&$(64, 50)$&$(G_2)_8^{} \times SU(2)_{24}^2$\\  \hline
14&$\begin{matrix} [2^3] \\ [4,1^2]\end{matrix}\quad 0$&$(0, 3, 1)$&$(45, 32)$&$(G_2)_8^{} \times SU(2)_{24}^{} \times SU(2)_5^{}$\\  \hline
15&$\begin{matrix} [1^6] \\ [1^6]\end{matrix}\quad A_1$&$(0, 5, 4)$&$(102, 79)$&$Sp(3)_8^2 \times SU(2)_{14}$\\  \hline
16&$\begin{matrix} [1^6] \\ [2,1^4]\end{matrix}\quad A_1$&$(0, 4, 4)$&$(92, 72)$&$Sp(2)_7^{} \times Sp(3)_8^{} \times SU(2)_{14}$\\  \hline
17&$\begin{matrix} [1^6] \\ [2^2,1^2]\end{matrix}\quad A_1$&$(1, 4, 3)$&$(84, 66)$&$Sp(3)_8^{} \times SU(2)_{14} \times SU(2)_6^{} \times U(1)^{}$\\  \hline
18&$\begin{matrix} [1^6] \\ [2^3]\end{matrix}\quad A_1$&$(0, 4, 3)$&$(78, 61)$&$Sp(3)_8^{} \times SU(2)_{14} \times SU(2)_{24}$\\  \hline
19&$\begin{matrix} [1^6] \\ [4,1^2]\end{matrix}\quad A_1$&$(0, 3, 2)$&$(59, 43)$&$Sp(3)_8^{} \times SU(2)_{14} \times SU(2)_5^{}$\\  \hline
20&$\begin{matrix} [2,1^4] \\ [2,1^4]\end{matrix}\quad A_1$&$(0, 3, 4)$&$(82, 65)$&$Sp(2)_7^2 \times SU(2)_{14}$\\  \hline
21&$\begin{matrix} [2,1^4] \\ [2^2,1^2]\end{matrix}\quad A_1$&$(1, 3, 3)$&$(74, 59)$&$Sp(2)_7^{} \times SU(2)_{14} \times SU(2)_6^{} \times U(1)^{}$\\  \hline
22&$\begin{matrix} [2,1^4] \\ [2^3]\end{matrix}\quad A_1$&$(0, 3, 3)$&$(68, 54)$&$Sp(2)_7^{} \times SU(2)_{14} \times SU(2)_{24}$\\  \hline
23&$\begin{matrix} [2,1^4] \\ [4,1^2]\end{matrix}\quad A_1$&$(0, 2, 2)$&$(49, 36)$&$Sp(2)_7^{} \times SU(2)_{14} \times SU(2)_5^{}$\\  \hline
24&$\begin{matrix} [2^2,1^2] \\ [2^2,1^2]\end{matrix}\quad A_1$&$(2, 3, 2)$&$(66, 53)$&$SU(2)_{14} \times SU(2)_6^2 \times U(1)^2$\\  \hline
25&$\begin{matrix} [2^2,1^2] \\ [2^3]\end{matrix}\quad A_1$&$(1, 3, 2)$&$(60, 48)$&$SU(2)_{14} \times SU(2)_{24} \times SU(2)_6^{} \times U(1)^{}$\\  \hline
26&$\begin{matrix} [2^2,1^2] \\ [4,1^2]\end{matrix}\quad A_1$&$(1, 2, 1)$&$(41, 30)$&$SU(2)_{14} \times SU(2)_5^{} \times SU(2)_6^{} \times U(1)^{}$\\  \hline
27&$\begin{matrix} [2^3] \\ [2^3]\end{matrix}\quad A_1$&$(0, 3, 2)$&$(54, 43)$&$SU(2)_{14} \times SU(2)_{24}^2$\\  \hline
28&$\begin{matrix} [2^3] \\ [4,1^2]\end{matrix}\quad A_1$&$(0, 2, 1)$&$(35, 25)$&$SU(2)_{14} \times SU(2)_{24} \times SU(2)_5^{}$\\  \hline
\end{longtable}

\subsubsection{Mixed fixture}\label{mixed_fixture}
There is only one mixed fixture in the $([\beta], [\alpha], [\alpha])$ sector.
\begin{longtable}{|c|c|c|c|c|}
\hline
\#&Fixture&$(d_3,d_4,d_6)$&$(n_h, n_v)$&$G_{\text{global}}$\\
\hline 
1&$\begin{matrix} [4,1^2] \\ [4,1^2]\end{matrix}\quad A_1$&$(0, 1, 0)$&$(15, 7)$&$\frac{1}{2}(2)+[Sp(3)_5 \times SU(2)_8] \text{ SCFT}$\\
\hline
\captionsetup{width=15cm}
\caption{The only mixed fixture of $([\beta],[\alpha],[\alpha])$ sector.}
\end{longtable}

\subsubsection{Gauge theory fixtures}\label{gauge_theory_fixtures}
\noindent
Finally, the $([\beta], [\alpha], [\alpha])$ sector has 76 gauge theory fixtures, which we list below.

{
\scriptsize
\setlength\LTleft{-.25in}
\renewcommand{\arraystretch}{2}

\begin{longtable}{|c|c|c|c|c|c|}
\captionsetup{width=18cm}
\caption {Gauge theory fixtures of the $([\beta],[\alpha],[\alpha])$ sector.  $d_i$ are the graded Coulomb branch dimensions. The notation ``$+n$ free" in $G_\text{global}$ indicates an additional factor of $Sp(n)$ associated to $n$ free hypermultiplets.  $(n_h,n_v)$ are the effective number of hypers and vectors, \emph{after} subtracting the contribution of the free hypers (if present).}\label{gtfixtable}\\
\endfirsthead
\hline
\#&Fixture&$(d_2,d_3,d_4,d_6)$&$(n_h, n_v)$&$G_{\text{global}}$&Gauge Theory\\
\hline
\endhead
\endfoot
\endlastfoot
\hline
\#&Fixture&$(d_2,d_3,d_4,d_6)$&$(n_h, n_v)$&$G_{\text{global}}$&Gauge Theory\\
\hline
1&$\begin{matrix}[1^6]\\ [3^2]\end{matrix}\quad 0$&$(1, 0, 4, 2)$&$(72, 53)$&$(G_2)_8 \times Sp(3)_8 \times SU(2)_8$&$Sp(3) + [(E_7)_8] + [Sp(3)_8^2 \times SU(2)_8]$\\
\hline
2&$\begin{matrix}[1^6]\\ [4,2]\end{matrix}\quad 0$&$(1, 0, 3, 2)$&$(64, 46)$&$(G_2)_8 \times Sp(3)_8$&$Sp(3) + [(E_7)_8] + [Sp(6)_8]$\\
\hline
3&$\begin{matrix}[2,1^4]\\ [3^2]\end{matrix}\quad 0$&$(1, 0, 3, 2)$&$(62, 46)$&$(G_2)_8 \times Sp(2)_7 \times SU(2)_8$&$Sp(3) + [(E_7)_8] + [Sp(4)_8 \times Sp(2)_7]$\\
\hline
4&$\begin{matrix}[2,1^4]\\ [4,2]\end{matrix}\quad 0$&$(1, 0, 2, 2)$&$(54, 39)$&$(G_2)_8 \times Sp(2)_7$&$Sp(3) + [(E_7)_8] + [Sp(5)_7] + \frac{1}{2} (6)$\\
\hline
5&$\begin{matrix}[2^2,1^2]\\ [3^2]\end{matrix}\quad 0$&$(1,1, 3, 1)$&$(54, 40)$&$\begin{gathered}(G_2)_8 \times SU(2)_8\\ \times SU(2)_6 \times U(1)\end{gathered}$&$Sp(3) + [(E_7)_8] + [SU(8)_8\times SU(2)_6]$\\
\hline
6&$\begin{matrix}[2^2,1^2]\\ [4,2]\end{matrix}\quad 0$&$(1,1, 2, 1)$&$(46, 33)$&$(G_2)_8 \times SU(2)_6 \times U(1)$&$Sp(3) + [(E_6)_6] + [(E_7)_8] +(6)$\\
\hline
7&$\begin{matrix}[2^3]\\ [3^2]\end{matrix}\quad 0$&$(1,0, 3, 1)$&$(48, 35)$&$(G_2)_8 \times SU(2)_{24} \times SU(2)_8$&$Sp(3) + [(E_7)_8] + [(E_7)_8]$\\
\hline
8&$\begin{matrix}[2^3]\\ [4,2]\end{matrix}\quad 0$&$(1,0, 2, 1)$&$(40, 28)$&$(G_2)_8 \times SU(2)_{24}$&$Sp(3) + [(E_7)_8] + \frac{1}{2} (14') + \frac{3}{2}(6)$\\
\hline
9&$\begin{matrix}[3^2]\\ [3^2]\end{matrix}\quad 0$&$(2,0, 2, 0)$&$(32, 20)$&$SO(7)_8 \times SU(2)_8^2$&$Sp(2) \times SU(2)+ [(E_7)_8] + 2 (4,1)$\\
\hline
10&$\begin{matrix}[3^2]\\ [4, 1^2]\end{matrix}\quad 0$&$(1,0, 2, 0)$&$(29, 17)$&$SO(7)_8 \times SU(2)_8 \times SU(2)_5$&$Sp(2) + [(E_7)_8] + (5)$\\
\hline
11&$\begin{matrix}[3^2]\\ [4, 2]\end{matrix}\quad 0$&$(2,0, 1, 0)$&$(24, 13)$&$SO(8)_8 \times SU(2)_8$&$Sp(2) \times SU(2)+(4,2)+4(4,1)$\\
\hline
12&$\begin{matrix}[4,1^2]\\ [4, 2]\end{matrix}\quad 0$&$(1,0, 1, 0)$&$(21, 10)$&$SO(8)_8 \times SU(2)_5$&$Sp(2)+ (5) + 4(4)$\\
\hline
13&$\begin{matrix}[4,2]\\ [4, 2]\end{matrix}\quad 0$&$(2,0, 0, 0)$&$(16, 6)$&$SO(8)_4^2$&$SU(2) \times SU(2)+ 4(2,1) + 4(1,2)$\\
\hline
14&$\begin{matrix}[1^6]\\ [3^2]\end{matrix}\quad A_1$&$(1,0, 3, 2)$&$(62, 46)$&$Sp(3)_8 \times SU(2)_{14} \times SU(2)_8$&$Sp(3) + [Sp(3)_8^2 \times SU(2)_8] + (14)$\\
\hline
15&$\begin{matrix}[1^6]\\ [4,2]\end{matrix}\quad A_1$&$(1,0, 2, 2)$&$(54, 39)$&$Sp(3)_8\times SU(2)_{14}$&$Sp(3) + [Sp(6)_8] +(14)$\\
\hline
16&$\begin{matrix}[2,1^4]\\ [3^2]\end{matrix}\quad A_1$&$(1,0, 2, 2)$&$(52, 39)$&$SU(2)_{14} \times Sp(2)_7 \times SU(2)_8$&$Sp(3) + [Sp(4)_8 \times Sp(2)_7] + (14)$\\
\hline
17&$\begin{matrix}[2,1^4]\\ [4,2]\end{matrix}\quad A_1$&$(1,0, 1, 2)$&$(44, 32)$&$Sp(2)_7 \times SU(2)_{14}$&$Sp(3) + [Sp(5)_7] + (14) + \tfrac{1}{2} (6)$\\
\hline
18&$\begin{matrix}[2^2,1^2]\\ [3^2]\end{matrix}\quad A_1$&$(1,1, 2, 1)$&$(44, 33)$&$\begin{gathered}SU(2)_{14} \times SU(2)_8\\ \times SU(2)_6 \times U(1)\end{gathered}$&$Sp(3)+ [SU(8)_8\times SU(2)_6] + (14)$\\
\hline
19&$\begin{matrix}[2^2,1^2]\\ [4,2]\end{matrix}\quad A_1$&$(1,1, 1, 1)$&$(36, 26)$&$SU(2)_{14} \times SU(2)_6 \times U(1)$&$Sp(3) + [(E_6)_6] + (14) + (6)$\\
\hline
20&$\begin{matrix}[2^3]\\ [3^2]\end{matrix}\quad A_1$&$(1,0, 2, 1)$&$(38, 28)$&$SU(2)_{24} \times SU(2)_8  \times SU(2)_{14}$&$Sp(3) + [(E_7)_8] + (14)$\\
\hline
21&$\begin{matrix}[2^3]\\ [4,2]\end{matrix}\quad A_1$&$(1,0, 1, 1)$&$(30, 21)$&$SU(2)_{24}\times SU(2)_{14}$&$SU(2) + [(G_2)_8 \times SU(2)_{14}]$\\
\hline
22&$\begin{matrix}[3^2]\\ [3^2]\end{matrix}\quad A_1$&$(2,0, 1, 0)$&$(21, 13)$&$SU(2)_8^3 \times SU(2)_5+ 1\;\text{free}$&$\begin{aligned}Sp(2) \times SU(2) &+ 2 (4,1) + (4,2)\\& + (5,1) + (1,1)\end{aligned}$\\
\hline
23&$\begin{matrix}[4,1^2]\\ [3^2]\end{matrix}\quad A_1$&$(1,0, 1, 0)$&$(18, 10)$&$Sp(2)_{5}\times SU(2)_8^2+ 1\;\text{free}$&$Sp(2) + 2 (4) + 2 (5) + (1)$\\
\hline
24&$\begin{matrix}[4,2]\\ [3^2]\end{matrix}\quad A_1$&$(2,0, 0, 0)$&$(12, 6)$&$SU(2)_4^5+ 2\;\text{free}$&$\begin{aligned}SU(2)^2 &+ 2 (1,2) + 2 (2,1)\\& + (2,2) + 2 (1,1)\end{aligned}$\\
\hline
25&$\begin{matrix}[4,2]\\ [4, 1^2]\end{matrix}\quad A_1$&$(1,0, 0, 0)$&$(8, 3)$&$SO(8)_2+3\;\text{free}$&$SU(2) + 4 (2) + 3 (1)$\\
\hline
26&$\begin{matrix}[1^6]\\ [1^6]\end{matrix}\quad \widetilde{A}_1$&$(1,0,5,3)$&$(93,71)$&$Sp(3)_{8}^2\times SU(2)_5$&$\begin{aligned}Sp(2)&+[Sp(3)_{8}^2\times Sp(2)_8]\\&+(5)\end{aligned}$\\
\hline
27&$\begin{matrix}[1^6]\\ [2,1^4]\end{matrix}\quad \widetilde{A}_1$&$(1,0,4,3)$&$(83,64)$&$Sp(3)_{8} \times Sp(2)_7 \times SU(2)_5$&$\begin{aligned}Sp(2)&+[Sp(3)_{8}\times Sp(2)_8\times Sp(2)_7]\\&+(5)\end{aligned}$\\
\hline
28&$\begin{matrix}[1^6]\\ [2^2,1^2]\end{matrix}\quad \widetilde{A}_1$&$(1,1,4,2)$&$(75,58)$&$\begin{gathered}Sp(3)_{8}\times SU(2)_6\\ \times SU(2)_5\times U(1)\end{gathered}$&$\begin{aligned}Sp(2)&+[Sp(3)_{8}\times Sp(2)_8\times SU(2)_6\times U(1)]\\&+(5)\end{aligned}$\\
\hline
29&$\begin{matrix}[1^6]\\ [2^3]\end{matrix}\quad \widetilde{A}_1$&$(1,0,4,2)$&$(69,53)$&$Sp(3)_{8}\times SU(2)_{24} \times SU(2)_5$&$\begin{aligned}Sp(2)&+[Sp(3)_{8}\times Sp(2)_8\times SU(2)_{24}]\\&+(5)\end{aligned}$\\
\hline
30&$\begin{matrix}[1^6]\\ [3^2]\end{matrix}\quad \widetilde{A}_1$&$(2,0,3,1)$&$(53,38)$&$Sp(3)_8\times SU(2)_8\times SU(2)_5$&$\begin{aligned}Spin(8) \times Sp(2)&+3 (8_v, 1)+(8_{c/s},1)\\&+(1, 5)+\frac{1}{2}(8_{s/c},4)\end{aligned}$\\
\hline
31&$\begin{matrix}[1^6]\\ [4,1^2]\end{matrix}\quad \widetilde{A}_1$&$(1,0,3,1)$&$(50,35)$&$Sp(3)_{8}\times SU(2)_5^2$& $\begin{gathered}Sp(2)+[Sp(5)_{8}\times SU(2)_5]+(5)\\ \simeq Spin(7)+[Spin(7)_8\times SU(2)_5^2]+3(8)\end{gathered}$\\
\hline
32&$\begin{matrix}[1^6]\\ [4,2]\end{matrix}\quad \widetilde{A}_1$&$(2,0,2,1)$&$(45,31)$&$Sp(3)_8 \times SU(2)_5$&$\begin{aligned}Sp(2)\times Spin(7)&+(5,1)+3(1,8)+\frac{1}{2}(4,8)\end{aligned}$\\
\hline
33&$\begin{matrix}[2,1^4]\\ [2,1^4]\end{matrix}\quad \widetilde{A}_1$&$(1,0,3,3)$&$(73,57)$&$Sp(2)_{7}^2\times SU(2)_5$&$\begin{aligned}Sp(2)&+[Sp(2)_8\times Sp(2)_{7}^2]+(5)\end{aligned}$\\
\hline
34&$\begin{matrix}[2,1^4]\\ [2^2,1^2]\end{matrix}\quad \widetilde{A}_1$&$(1,1,3,2)$&$(65,51)$&$\begin{gathered}Sp(2)_{7} \times SU(2)_6\\ \times SU(2)_5 \times U(1)\end{gathered}$&$\begin{aligned}Sp(2)&+[Sp(2)_8\times Sp(2)_{7}^2\times SU(2)_6 \times U(1)]\\&+(5)\end{aligned}$\\
\hline
35&$\begin{matrix}[2,1^4]\\ [2^3]\end{matrix}\quad \widetilde{A}_1$&$(1,0,3,2)$&$(59,46)$&$Sp(2)_{7}\times SU(2)_{24} \times SU(2)_5$&$\begin{aligned}Sp(2)&+[Sp(2)_8\times Sp(2)_{7}\times SU(2)_{24} ]+(5)\end{aligned}$\\
\hline
36&$\begin{matrix}[2,1^4]\\ [3^2]\end{matrix}\quad \widetilde{A}_1$&$(2,0,2,1)$&$(43,31)$&$Sp(2)_7 \times SU(2)_8 \times SU(2)_5$&$\begin{aligned}Sp(2)&\times Spin(7)\\&+(5,1)+(1,8)+ \frac{1}{2}(4,8)+2(1,7)\end{aligned}$\\
\hline
37&$\begin{matrix}[2,1^4]\\ [4,1^2]\end{matrix}\quad \widetilde{A}_1$&$(1,0,2,1)$&$(40,28)$&$Sp(2)_7\times SU(2)_5^2$&$\begin{aligned}Sp(2)&+[Sp(4)_7\times SU(5)_{5}]+(5)+\frac{1}{2}(4)\end{aligned}$\\
\hline
38&$\begin{matrix}[2,1^4]\\ [4,2]\end{matrix}\quad \widetilde{A}_1$&$(2,0,1,1)$&$(35,24)$&$Sp(2)_7\times SU(2)_5$&$\begin{aligned}G_2\times Sp(2)&+2(7,1)+(1,5)\\&+\tfrac{1}{2}(7,4)+\tfrac{1}{2}(1,4)\end{aligned}$\\
\hline
39&$\begin{matrix}[2^2,1^2]\\ [2^2,1^2]\end{matrix}\quad \widetilde{A}_1$&$(1,2,3,1)$&$(57,45)$&$SU(2)_6^2\times SU(2)_5 \times U(1)^2$&$\begin{aligned}Sp(2)&+[Sp(2)_8 \times SU(2)_6^2 \times U(1)^2]+(5)\end{aligned}$\\
\hline
40&$\begin{matrix}[2^2,1^2]\\ [2^3]\end{matrix}\quad \widetilde{A}_1$&$(1,1,3,1)$&$(51,40)$&$\begin{gathered}SU(2)_{24} \times SU(2)_6\\ \times SU(2)_5 \times U(1)\end{gathered}$&$\begin{aligned}Sp(2)&+[Sp(2)_8\times SU(2)_{24}\times SU(2)_6 \times U(1)]\\&+(5)\end{aligned}$\\
\hline
41&$\begin{matrix}[2^2,1^2]\\ [3^2]\end{matrix}\quad \widetilde{A}_1$&$(2,1,2,0)$&$(35,25)$&$\begin{gathered}SU(2)_8\times SU(2)_6\\ \times SU(2)_5\times U(1)^2\end{gathered}$&$\begin{aligned}SU(4)&\times Sp(2)+(4,4)+2(4,1)\\&+(6,1)+(1,5)\end{aligned}$\\
\hline
42&$\begin{matrix}[2^2,1^2]\\ [4,1^2]\end{matrix}\quad \widetilde{A}_1$&$(1,1,2,0)$&$(32,22)$&$SU(2)_6 \times SU(2)_5^2 \times U(1)^2$&$\begin{aligned}Sp(2)&+[Sp(3)_6\times SU(2)_5 \times U(1)]+(5)+(4)\end{aligned}$\\
\hline
43&$\begin{matrix}[2^2,1^2]\\ [4,2]\end{matrix}\quad \widetilde{A}_1$&$(2,1,1,0)$&$(27,18)$&$SU(2)_6 \times SU(2)_5 \times U(1)^3$&$\begin{aligned}Sp(2)&\times SU(3)\\&+(5,1)+2(1,3)+ (4,3) + (4,1)\end{aligned}$\\
\hline
44&$\begin{matrix}[2^3]\\ [2^3]\end{matrix}\quad \widetilde{A}_1$&$(1,0,3,1)$&$(45,35)$&$SU(2)_{24}^2\times SU(2)_5$&$\begin{aligned}Sp(2)&+[Sp(2)_8\times SU(2)_{24}^2]+(5)\end{aligned}$\\
\hline
45&$\begin{matrix}[2^3]\\ [3^2]\end{matrix}\quad \widetilde{A}_1$&$(2,0,2,0)$&$(29,20)$&$SU(2)_{16}\times SU(2)_8^2\times SU(2)_5$&$\begin{aligned}Sp(2)^2&+(4,4)+2(1,4)+(5,1)\end{aligned}$\\
\hline
46&$\begin{matrix}[2^3]\\ [4,1^2]\end{matrix}\quad \widetilde{A}_1$&$(1,0,2,0)$&$(26,17)$&$SU(2)_{16}\times SU(2)_8\times SU(2)_5^2$&$\begin{aligned}Sp(2)&+[Sp(3)_5\times SU(2)_8]\\&+\tfrac{3}{2}(4)+(5)\end{aligned}$\\
\hline
47&$\begin{matrix}[2^3]\\ [4,2]\end{matrix}\quad \widetilde{A}_1$&$(2,0,1,0)$&$(21,13)$&$SU(2)_8^3 \times SU(2)_5$&$\begin{aligned}Sp(2)\times SU(2)&+\tfrac{1}{2}(5,2)\\&+\tfrac{3}{2}(1,2)+(5,1)+2(4,1)\end{aligned}$\\
\hline
48&$\begin{matrix}[1^6]\\ [1^6]\end{matrix}\quad G_2(a_1)$&$(2,0,4,3)$&$(88,67)$&$Sp(3)_8^2$&$\begin{aligned}Spin(8)^2&+[(E_8)_{12}]\\&+3(8_v,1)+3(1,8_v)\end{aligned}$\\
\hline
49&$\begin{matrix}[1^6]\\ [2,1^4]\end{matrix}\quad G_2(a_1)$&$(2,0,3,3)$&$(78,60)$&$Sp(3)_8\times Sp(2)_7$&$\begin{aligned}Spin(8)\times Spin(7)&+[(E_8)_{12}]\\&+3(8_v,1)+2(1,7)\end{aligned}$\\
\hline
50&$\begin{matrix}[1^6]\\ [2^2,1^2]\end{matrix}\quad G_2(a_1)$&$(2,1,3,2)$&$(70,54)$&$Sp(3)_8\times SU(2)_{6}\times U(1)$&$\begin{aligned}Spin(8)\times SU(4)&+[(E_8)_{12}]\\&+3(8_v,1)+(1,6)\end{aligned}$\\
\hline
51&$\begin{matrix}[1^6]\\ [2^3]\end{matrix}\quad G_2(a_1)$&$(2,0,3,2)$&$(64,49)$&$Sp(3)_8\times SU(2)_{24}$&$\begin{aligned}Spin(8)\times Sp(2)&+[(E_8)_{12}]+3(8_v,1)\end{aligned}$\\
\hline
52&$\begin{matrix}[1^6]\\ [3^2]\end{matrix}\quad G_2(a_1)$&$(3,0,2,1)$&$(48,34)$&$Sp(3)_8\times SU(2)_8$&$\begin{aligned}Spin(8)\times SU(2)^2&+3(8_v,1,1)+(8_s,1,1)\\&+\tfrac{1}{2}(8_s,2,1)+\tfrac{1}{2}(8_c,1,2)\end{aligned}$\\
\hline
53&$\begin{matrix}[1^6]\\ [4,1^2]\end{matrix}\quad G_2(a_1)$&$(2,0,2,1)$&$(45,31)$&$Sp(3)_8\times SU(2)_5$&$\begin{aligned}Spin(7)\times Sp(2)&+3(8,1)\\&+(1,5)+\tfrac{1}{2}(8,4)\end{aligned}$\\
\hline
54&$\begin{matrix}[1^6]\\ [4,2]\end{matrix}\quad G_2(a_1)$&$(3,0,1,1)$&$(40,27)$&$Sp(3)_8$&$\begin{aligned}Spin(7)\times SU(2)^2&+3(8,1,1)\\&+\tfrac{1}{2}(8,2,1)+\tfrac{1}{2}(8,1,2)\end{aligned}$\\
\hline
55&$\begin{matrix}[2,1^4]\\ [2,1^4]\end{matrix}\quad G_2(a_1)$&$(2,0,2,3)$&$(68,53)$&$Sp(2)_7^2$&$\begin{aligned}Spin(7)^2+[(E_8)_{12}]+2(7,1)+2(1,7)\end{aligned}$\\
\hline
56&$\begin{matrix}[2,1^4]\\ [2^2,1^2]\end{matrix}\quad G_2(a_1)$&$(2,1,2,2)$&$(60,47)$&$Sp(2)_7\times SU(2)_6\times U(1)$&$\begin{aligned}Spin(7)\times SU(4)&+[(E_8)_{12}]\\&+2(7,1)+(1,6)\end{aligned}$\\
\hline
57&$\begin{matrix}[2,1^4]\\ [2^3]\end{matrix}\quad G_2(a_1)$&$(2,0,2,2)$&$(54,42)$&$Sp(2)_7\times SU(2)_{24}$&$\begin{aligned}Spin(7)\times Sp(2)+[(E_8)_{12}]+2(7,1)\end{aligned}$\\
\hline
58&$\begin{matrix}[2,1^4]\\ [3^2]\end{matrix}\quad G_2(a_1)$&$(3,0,1,1)$&$(38,27)$&$Sp(2)_7\times SU(2)_8$&$\begin{aligned}Spin(7)\times SU(2)^2&+2(7,1,1)+(8,1,1)\\&+\tfrac{1}{2}(8,2,1)+\tfrac{1}{2}(8,1,2)\end{aligned}$\\
\hline
59&$\begin{matrix}[2,1^4]\\ [4,1^2]\end{matrix}\quad G_2(a_1)$&$(2,0,1,1)$&$(35, 24)$&$Sp(2)_7\times SU(2)_5$&$\begin{aligned}G_2\times Sp(2)&+2(7,1)+(1,5)\\&+\tfrac{1}{2}(7,4)+\tfrac{1}{2}(1,4)\end{aligned}$\\
\hline
60&$\begin{matrix}[2,1^4]\\ [4,2]\end{matrix}\quad G_2(a_1)$&$(3,0,0,1)$&$(30, 20)$&$Sp(2)_7$&$\begin{aligned}G_2\times SU(2)^2&+2(7,1,1)\\&+\tfrac{1}{2}(7,2,1)+\tfrac{1}{2}(7,1,2)\\&+\tfrac{1}{2}(1,2,1)+\tfrac{1}{2}(1,1,2) \end{aligned}$\\
\hline
61&$\begin{matrix}[2^2,1^2]\\ [2^2,1^2]\end{matrix}\quad G_2(a_1)$&$(2,2,2,1)$&$(52,41)$&$SU(2)_{6}^2\times U(1)^2$&$\begin{aligned}SU(4)^2+[(E_8)_{12}]+(6,1)+(1,6)\end{aligned}$\\
\hline
62&$\begin{matrix}[2^2,1^2]\\ [2^3]\end{matrix}\quad G_2(a_1)$&$(2, 1, 2, 1)$&$(46,36)$&$SU(2)_{24}\times SU(2)_{6}\times U(1)$&$\begin{aligned}SU(4)\times Sp(2)+[(E_8)_{12}]+(6,1)\end{aligned}$\\
\hline
63&$\begin{matrix}[2^2,1^2]\\ [3^2]\end{matrix}\quad G_2(a_1)$&$(3, 1, 1, 0)$&$(30,21)$&$SU(2)_8\times SU(2)_{6}\times U(1)^3$&$\begin{aligned}SU(4)\times SU(2)^2 &+2(4,1,1)+(6,1,1)\\&+(4,2,1)+(4,1,2)\end{aligned}$\\
\hline
64&$\begin{matrix}[2^2,1^2]\\ [4,1^2]\end{matrix}\quad G_2(a_1)$&$(2, 1, 1, 0)$&$(27,18)$&$SU(2)_{6}\times SU(2)_5\times U(1)^3$&$\begin{aligned}SU(3)\times Sp(2) &+2(3,1)+(3,4)\\&+(1,5)+(1,4)\end{aligned}$\\
\hline
65&$\begin{matrix}[2^2,1^2]\\ [4,2]\end{matrix}\quad G_2(a_1)$&$(3, 1, 0, 0)$&$(22,14)$&$SU(2)_{6}\times U(1)^5$&$\begin{aligned}SU(3)\times SU(2)^2 &+2(3,1,1)\\&+(3,2,1)+(3,1,2)\\&+(1,2,1)+(1,1,2)\end{aligned}$\\
\hline
66&$\begin{matrix}[2^3]\\ [2^3]\end{matrix}\quad G_2(a_1)$&$(2, 0, 2, 1)$&$(40,31)$&$SU(2)_{24}^2$&$\begin{aligned}Sp(3)\times Sp(2)&+\tfrac{3}{2}(6,1)+\tfrac{1}{2}(14',1)\\&+(6,4)\end{aligned}$\\
\hline
67&$\begin{matrix}[2^3]\\ [3^2]\end{matrix}\quad G_2(a_1)$&$(3, 0, 1, 0)$&$(24,16)$&$SU(2)_8^4$&$\begin{aligned}Sp(2)\times SU(2)^2&+2(4,1,1)\\&+(4,2,1)+(4,1,2)\end{aligned}$\\
\hline
68&$\begin{matrix}[2^3]\\ [4,1^2]\end{matrix}\quad G_2(a_1)$&$(2, 0, 1, 0)$&$(21,13)$&$SU(2)_8^3\times SU(2)_5$&$\begin{aligned}Sp(2)\times SU(2)&+2(4,1)+(4,2)\\&+(5,1)\end{aligned}$\\
\hline
69&$\begin{matrix}[2^3]\\ [4,2]\end{matrix}\quad G_2(a_1)$&$(3, 0, 0, 0)$&$(16,9)$&$SU(2)_4^6$&$\begin{aligned}SU(2)^3&+2(2,1,1)+2(1,2,1)\\&+2(1,1,2)+\tfrac{1}{2}(2,2,2)\end{aligned}$\\
\hline
70&$\begin{matrix}[1^6]\\ [1^6]\end{matrix}\quad G_2$&$(1, 0, 2, 1)$&$(48,28)$&$Sp(3)_8^2$&$Spin(8)+3(8_s)+3(8_v)$\\
\hline
71&$\begin{matrix}[1^6]\\ [2,1^4]\end{matrix}\quad G_2$&$(1, 0, 1, 1)$&$(38,21)$&$Sp(3)_8\times Sp(2)_7$&$Spin(7)+3(8)+2(7)$\\
\hline
72&$\begin{matrix}[1^6]\\ [2^2,1^2]\end{matrix}\quad G_2$&$(1, 1, 1, 0)$&$(30,15)$&$SU(6)_8\times SU(2)_6\times U(1)$&$SU(4)+6(4)+1(6)$\\
\hline
73&$\begin{matrix}[1^6]\\ [2^3]\end{matrix}\quad G_2$&$(1, 0, 1, 0)$&$(24,10)$&$SO(12)_4$&$Sp(2)+6(4)$\\
\hline
74&$\begin{matrix}[2,1^4]\\ [2,1^4]\end{matrix}\quad G_2$&$(1, 0, 0, 1)$&$(28,14)$&$Sp(4)_7$&$G_2 +4(7)$\\
\hline
75&$\begin{matrix}[2,1^4]\\ [2^2,1^2]\end{matrix}\quad G_2$&$(1, 1, 0, 0)$&$(18, 8)$&$SU(6)_6\times U(1)+2\;\text{free}$&$SU(3) + 6(3) + 2(1)$\\
\hline
76&$\begin{matrix}[2,1^4]\\ [2^3]\end{matrix}\quad G_2$&$(1, 0, 0, 0)$&$(8, 3)$&$SO(8)_2 + 6\;\text{free}$&$SU(2) + 4 (2) + 6 (1)$\\
\hline
\end{longtable}
}

\section{4-punctured sphere with 4 $[\alpha]$-twisted punctures}\label{4punctured_sphere_with_4_twisted_punctures}

As a preliminary to \S\ref{resolving_atypical_punctures}, let us make some remarks on the moduli space of $S_3$ bundles on a 4-punctured sphere, with 4 punctures in the $[\alpha]$ conjugacy class. We can obtain the 4-punctured sphere by gluing together two of our 3-punctured spheres \eqref{unzipped} along $\gamma_3=(\tilde{\gamma_3})^{-1}$, yielding the following figure.

\begin{equation}
\begin{matrix}
 \includegraphics[width=220pt]{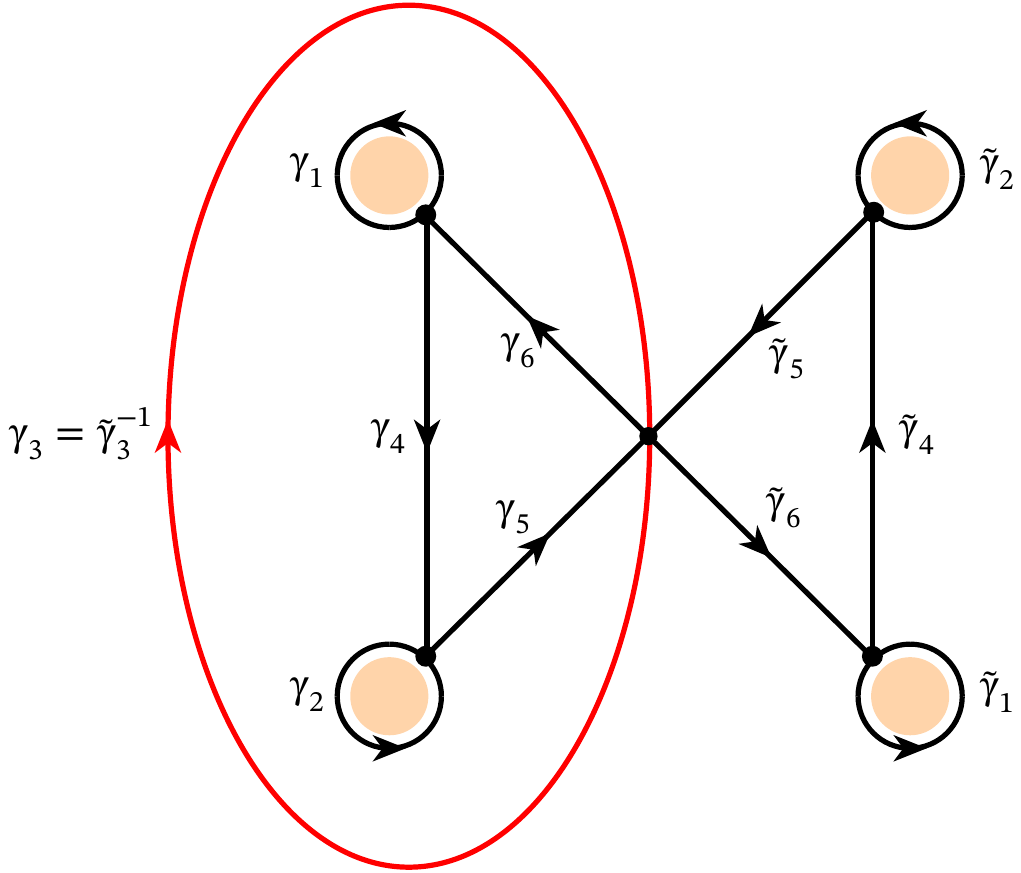}
 \end{matrix}
\label{3ptCWsewn}\end{equation}
The isomorphism classes of $S_3$ bundles on a 4-punctured sphere are labeled by the holonomies around a basis for $\pi_1$ of the punctured surface. In this case, we can choose the basepoint to be the distinguished point on $\gamma_3$ and --- for present purposes --- forget about $\gamma_3$ (and $\gamma_4,\tilde{\gamma}_4$).

\begin{displaymath}
 \includegraphics[width=161pt]{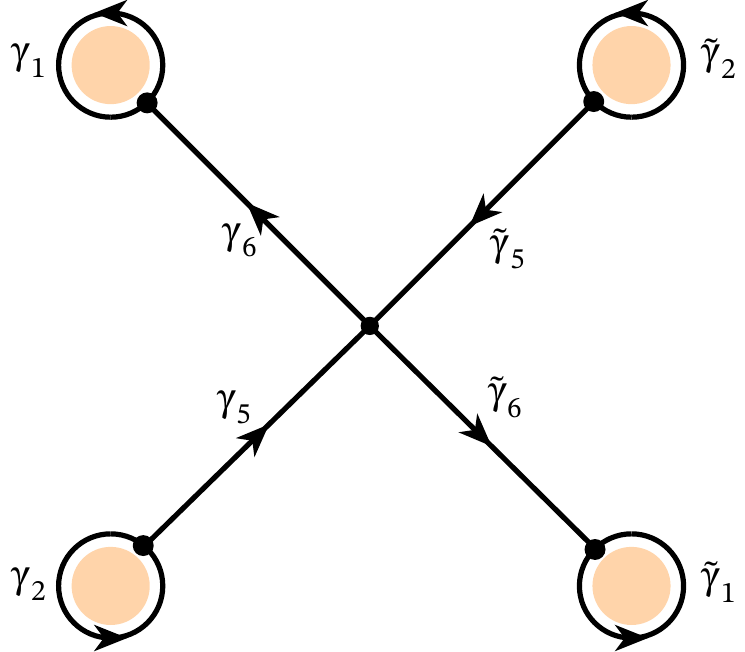}
\end{displaymath}
The holonomies are

\begin{equation}
\delta_1= \gamma_6^{-1}\gamma_1\gamma_6,\quad\delta_2=\gamma_5\gamma_2\gamma_5^{-1},\quad \delta_3=\tilde{\gamma}_6^{-1}\tilde{\gamma}_1\tilde{\gamma}_6,\quad \delta_4=\tilde{\gamma}_5\tilde{\gamma}_2\tilde{\gamma}_5^{-1}
\label{deltas}\end{equation}
which satisfy one relation,

\begin{displaymath}
\delta_4\delta_3\delta_2\delta_1=\mathbb{1}
\end{displaymath}
If the four punctures are in the $[\alpha]$ conjugacy class ($[\gamma_1]=[\gamma_2]=[\tilde{\gamma}_1]=[\tilde{\gamma}_2]=[\alpha])$, then it is easy to show \cite{Distler:toappear} that

\begin{itemize}%
\item Either all four are equal $\delta_1=\delta_2=\delta_3=\delta_4=\alpha \beta^j$ (for some choice of $j$) or two of them are equal.
\item When all four are equal, modular transformations leave the $\delta$s invariant, and the moduli space of $S_3$ bundles is a copy of $\overline{\mathcal{M}}_{0,4}$. In that case, the twist along the original cycle along which we glued, $\gamma_3$, is trivial. Indeed the twist around \emph{any} pinching cycle, when the 4-punctured sphere degenerates, is $\mathbb{1}$.
\item When they are not all equal, the consistent choices of $\delta$s form a single orbit under the modular group. In this case, the twists around the pinching cycle could be any of $\mathbb{1},\beta$ or $\beta^2$ and all three possibilities appear for each degeneration of the 4-punctured sphere.
\item In either case, the physics is invariant under an overall conjugation of the $\delta_i$ by a common element of $S_3$.

\end{itemize}

The conformal manifold in the former case is a copy of $\overline{\mathcal{M}}_{0,4}$ or a quotient thereof (if some of the nilpotent orbits are identical). In the latter case, the conformal manifold is an 4-fold branched cover\footnote{Choosing 4 elements of the $[\alpha]$ conjugacy class, which are not all equal and whose product is $\mathbb{1}$, give 24 possibilities. An overall conjugation reduces this number to 4.} of $\overline{\mathcal{M}}_{0,4}$, whose interior is  $\text{UHP}/\Gamma(4)$, where $\Gamma(4)\subset\Gamma(2)$ is an index-8 normal subgroup
\begin{equation*}
1\to \Gamma(4)\to \Gamma(2)\to (\mathbb{Z})_2^3\to 1
\end{equation*}
One of the $(\mathbb{Z}_2)$s is the center, generated by $-\mathrm{I}$, which acts trivially on the UHP. The six cusp points of $\Gamma(4)$ map $2:1$ to the boundary points of $\overline{\mathcal{M}}_{0,4}$. Correspondingly, in each degeneration limit, the four-punctured sphere may obtain up to two weakly-coupled descriptions based on whether the twist around the degenerating cycle is $\mathbb{1}$ or in the $[\beta]$ conjugacy class. This gives new conformal manifolds, not previously seen in the literature that may have more than 3 and up to 6 different weakly coupled limits. We will see instances of this result below. A more systematic discussion will be presented in \cite{Distler:toappear}.

\subsection{Resolving atypical punctures}\label{resolving_atypical_punctures}

The $\mathbb{Z}_3$ twisted sector has two atypical punctures, $\tilde{A}_1$ and $G_2$, which support a ``hidden'' exactly marginal deformation. They resolve to a pair of punctures from non-commuting $\mathbb{Z}_2$-twisted sectors: $\tilde{A}_1 \sim [4,1^2]+ [6]$ and $G_2 \sim [6]+[6]$. This \emph{appears} to lead to choices (of which $[\alpha]$-twisted sectors the resolving punctures come from). In fact, all of these choices lead to isomorphic families of SCFTs, as we shall now explain.

Consider, as an example, the gauge-theory fixture

\begin{displaymath}
 \includegraphics[width=88pt]{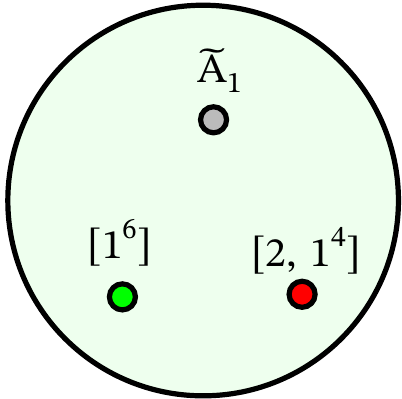}
\end{displaymath}
where, in this case, we can choose the internal twists to be trivial. ``Resolving'' the $\tilde{A}_1$ puncture means replacing this with a 4-punctured sphere in the degeneration limit where a $[4,1^2]$ and a $[6]$ puncture (from \emph{different} $\mathbb{Z}_2$-twisted sectors) collide.

It is always possible to choose a representative so that the internal twists remain trivial, in which case, we have three possible resolutions.

\begin{equation}
\begin{gathered}
 \includegraphics[width=190pt]{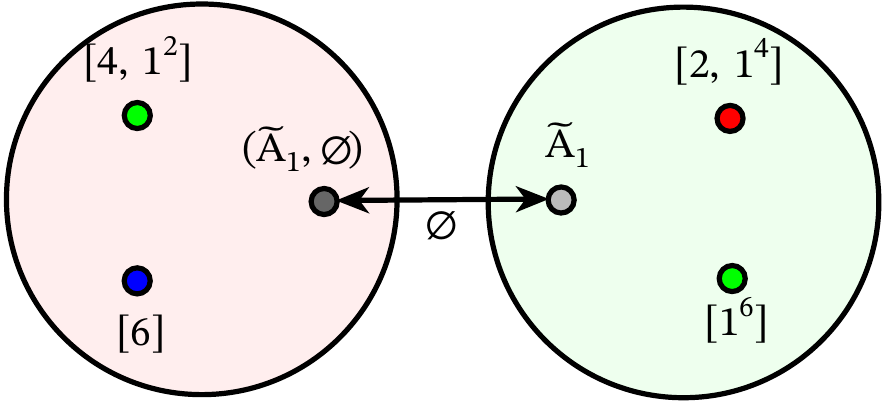}\\
 \includegraphics[width=190pt]{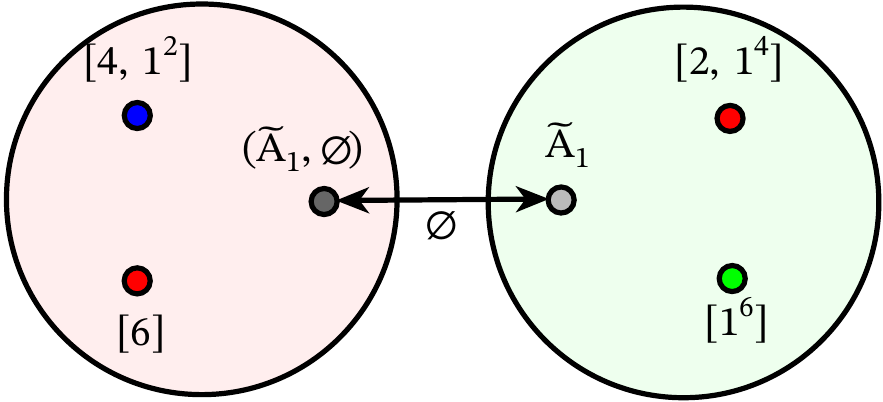}\\
 \includegraphics[width=190pt]{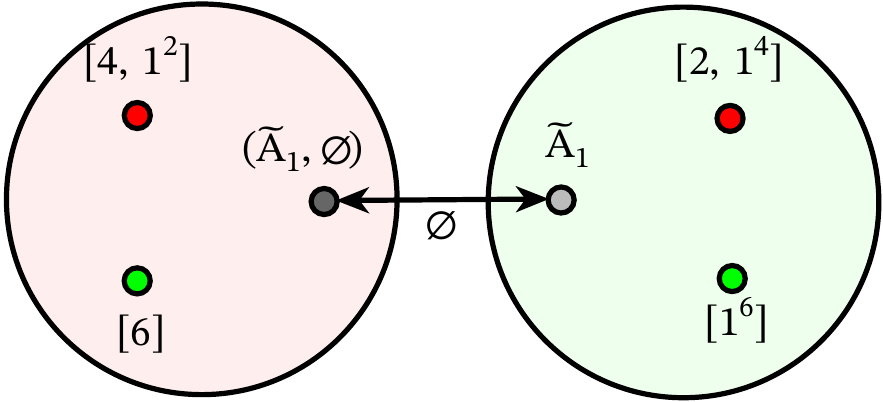}
\end{gathered}
\label{DifferentResolutions}\end{equation}
Since the internal twists are trivial, the $\delta$s defined in \eqref{deltas} are just equal to the corresponding external twists. Since these are not all equal, we are in the situation where the conformal manifold is a branched cover of $\overline{\mathcal{M}}_{0,4}$. Over this point on the boundary, the twist around the pinching cycle can be $\mathbb{1},\beta$ or $\beta^2$. If it is $\beta$ or $\beta^2$, we have the gauge theory fixture depicted above. If it is $\mathbb{1}$, then we have an $Sp(2)$ gauge theory\footnote{We use different colors to specify different internal twists. The labeling is the same as for external twists that we mentioned in footnote \ref{ft:colorpunctures}: white represents a trivial twist; green, red and blue represent $\alpha$, $\alpha \beta$ and $\alpha \beta^2$ twists; and light grey and dark grey represent $\beta$ and $\beta^2$ twists respectively. }.

\begin{equation}\label{IIBsp2gauged}
\begin{matrix}
 \includegraphics[width=190pt]{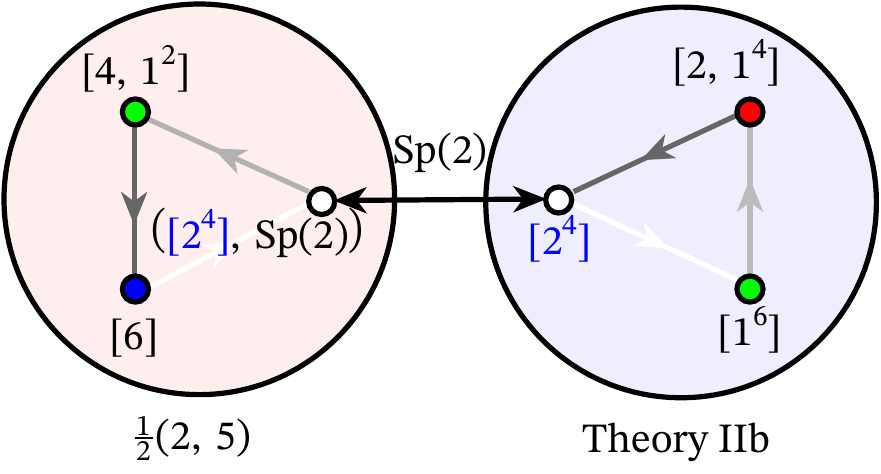}
 \end{matrix}
\end{equation}
The fixture on the left is a hypermultiplet in the $(5)$ of $Sp(2)$. The fixture on the right is one of the two SCFTs with flavor symmetry algebra $Sp(3)_8 \times Sp(2)_8 \times Sp(2)_7$ recently studied in \cite{Distler:2020tub}. The internal twists on the fixture determine it to be Theory IIb in the notation of that paper.

Now consider the degeneration limit where $[4,1^2]$ collides with the $[2,1^4]$ puncture. The twist around the degenerating cycle can be $\mathbb{1}$, $\beta$ or $\beta^2$. When it is $\mathbb{1}$, we obtain a $Spin(8)$ gauge theory,

\begin{displaymath}
 \includegraphics[width=216pt]{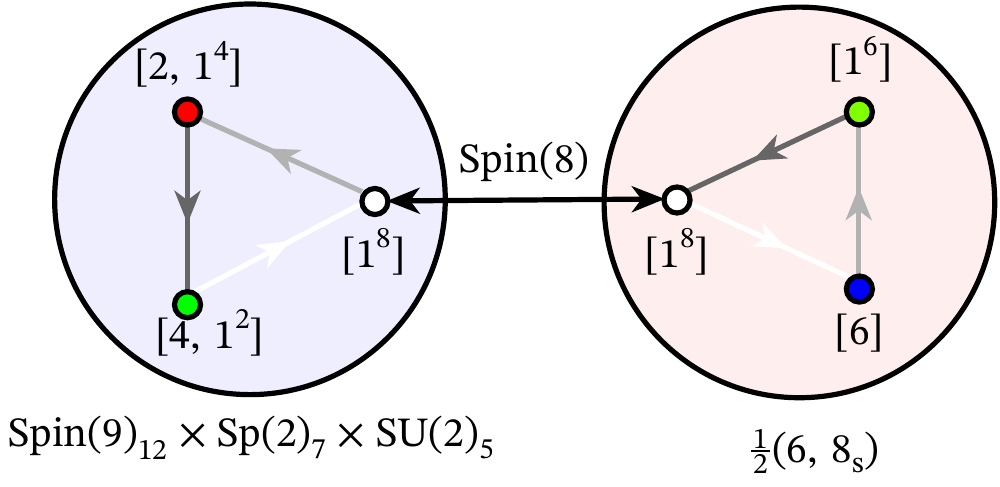}
\end{displaymath}
and when it is $\beta$ or $\beta^2$, we obtain a $G_2$ gauge theory.
\begin{displaymath}
 \includegraphics[width=190pt]{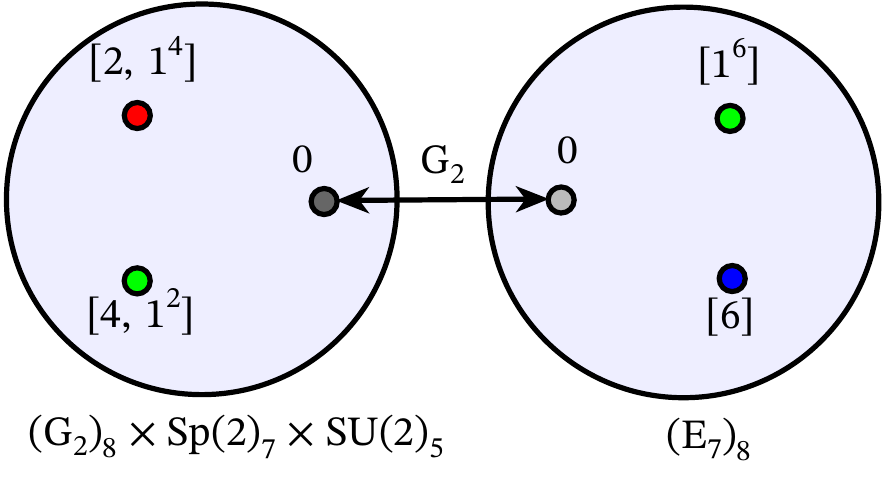}
\end{displaymath}

Finally, there is the degeneration where  $[4,1^2]$ collides with $[1^6]$. When the twist around the degenerating cycle is $\mathbb{1}$, we obtain a $Spin(7)$ gauge theory
\begin{displaymath}
 \includegraphics[width=204pt]{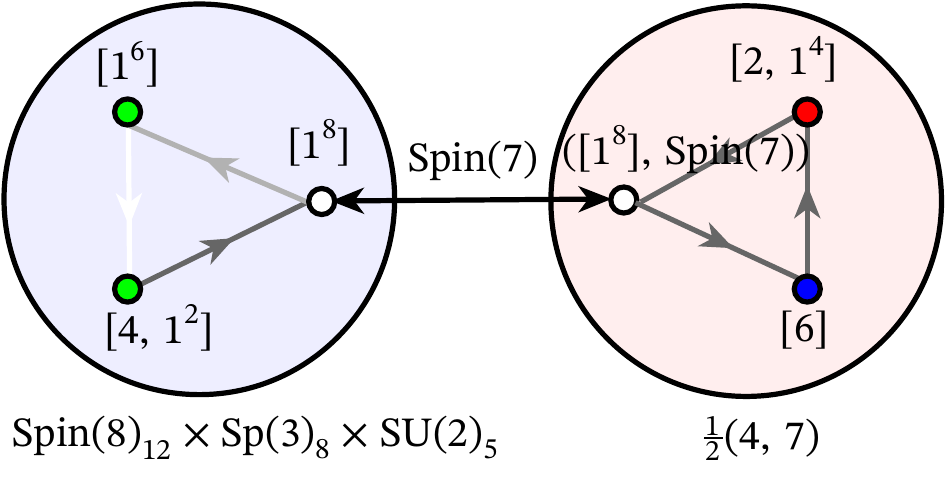}
\end{displaymath}
and when it is $\beta$ or $\beta^2$ we obtain a $G_2$ gauge theory.
\begin{displaymath}
 \includegraphics[width=190pt]{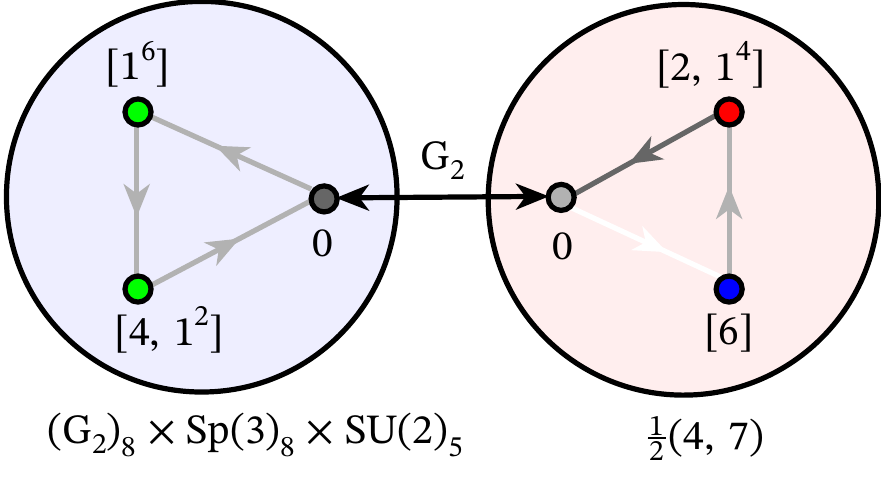}
\end{displaymath}
Overall, this family of SCFTs has \emph{five} distinct weakly-coupled limits. This is a consequence of the nonabelian nature of $S_3$ twisted $D_4$ sector which has not been seen previously. Indeed, all previously known one-dimensional conformal manifolds of 4D $\mathcal{N}=2$ SCFTs have at most three distinct weakly-coupled points.

Let us compare this with the family of SCFTs obtained from the four-punctured sphere in the second row of \eqref{DifferentResolutions}. When the twist around the degenerating cycle is $\mathbb{1}$, we obtain the same $Sp(2)$ gauge theory as above.

\begin{displaymath}
 \includegraphics[width=190pt]{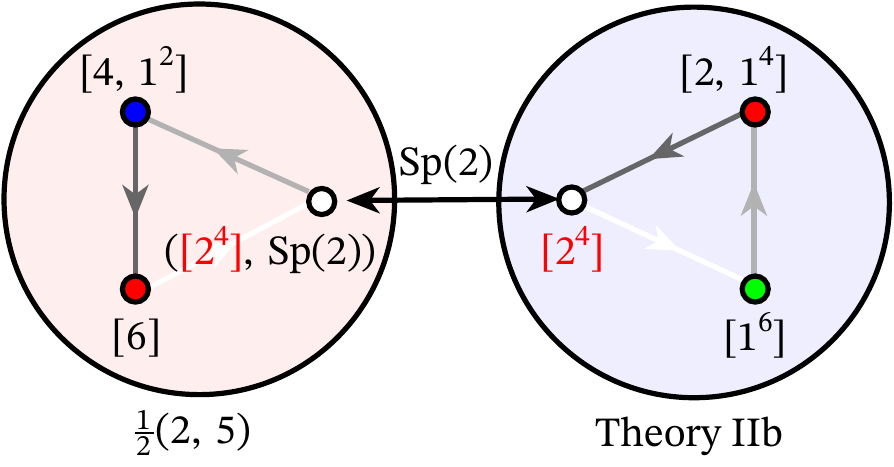}
\end{displaymath}
The weakly coupled descriptions also match in the other degeneration limits. When $[4,1^2]$ collides with the $[2,1^4]$ puncture, we obtain the same $Spin(8)$ and $G_2$ gauge theories,

\begin{displaymath}
 \includegraphics[width=204pt]{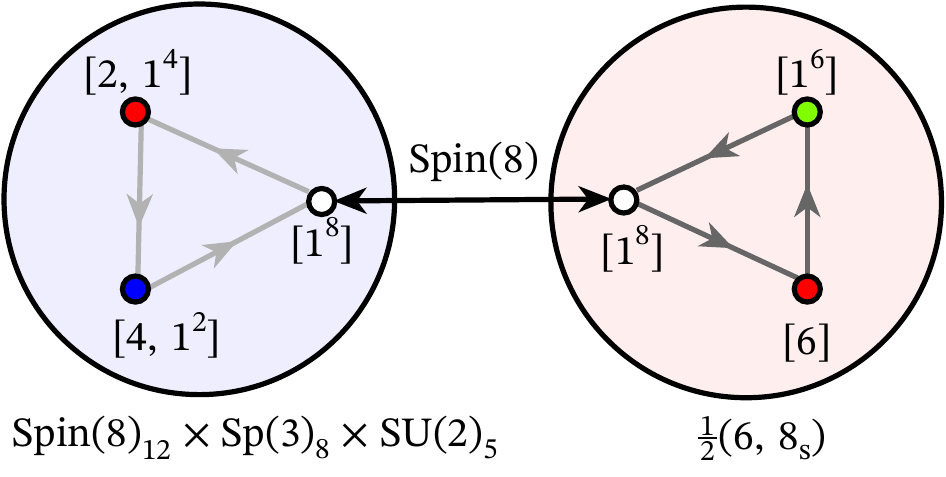}
\end{displaymath}
\begin{displaymath}
 \includegraphics[width=190pt]{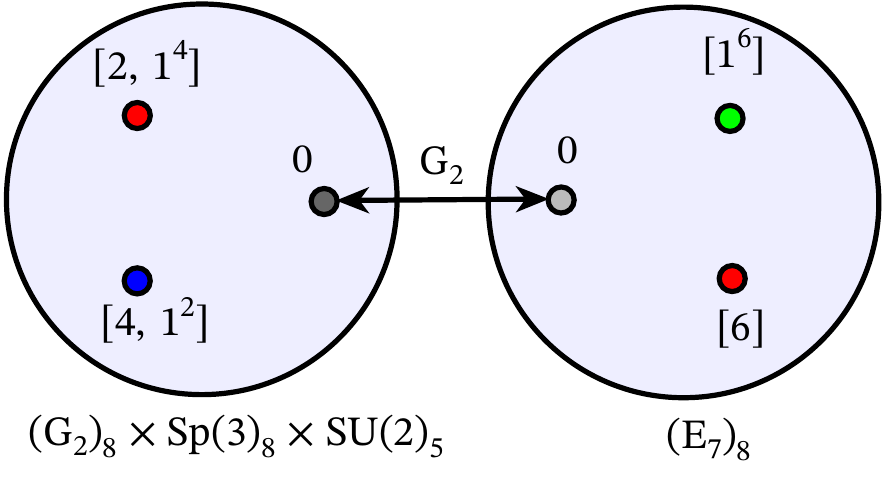}
\end{displaymath}
and when $[4,1^2]$ collides with the $[1^6]$ puncture, we obtain the same $Spin(7)$ and $G_2$ gauge theories.

\begin{displaymath}
 \includegraphics[width=204pt]{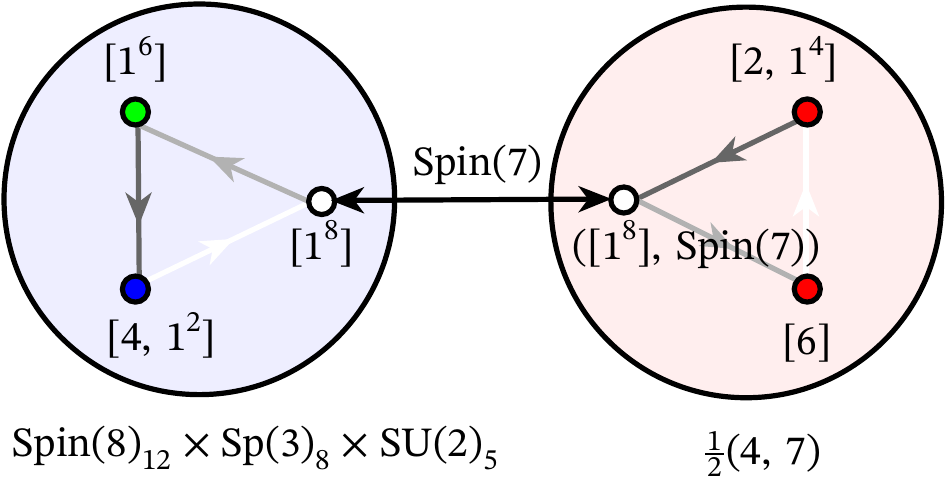}
\end{displaymath}
\begin{displaymath}
 \includegraphics[width=190pt]{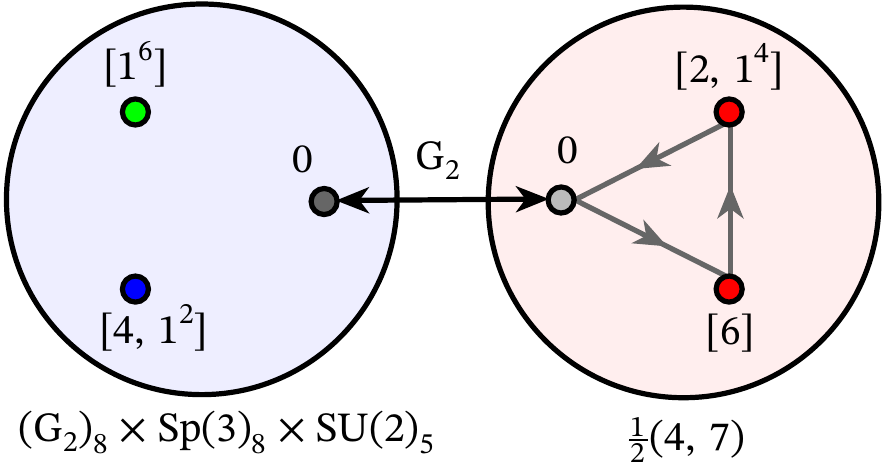}
\end{displaymath}
It is straightforward to check that the third row in \eqref{DifferentResolutions} yields the same family of SCFTs.

By way of contrast, we could have started with the same 4-punctured sphere obtained by resolving the $\tilde{A}_1$ puncture, but with the internal twists chosen so that the $\delta$s are all equal. For concreteness, take the first row of \eqref{DifferentResolutions} but let $\delta_1 = \delta_2 = \delta_3 = \delta_4 = \alpha$.

\begin{equation}\label{IIAgauged}
\begin{matrix}
 \includegraphics[width=190pt]{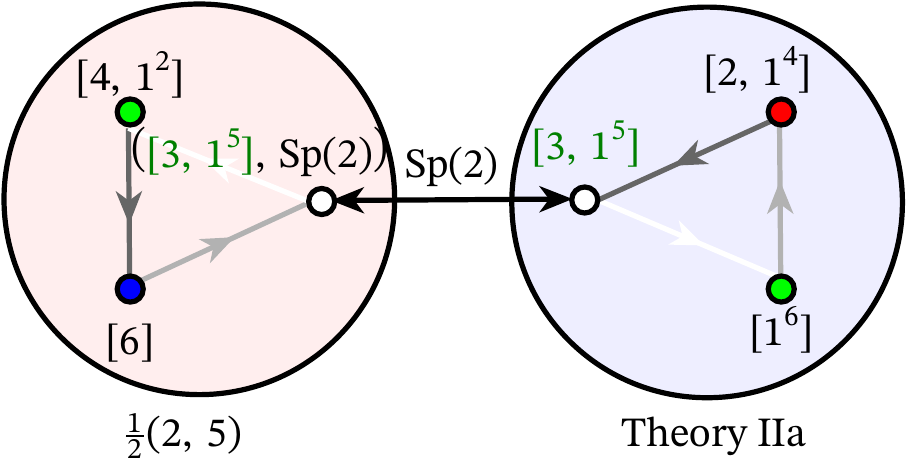}
 \end{matrix}
\end{equation}
The fixture on the right has the same flavor symmetry algebra as above, i.e. $Sp(3)_8 \times Sp(2)_8 \times Sp(2)_7$ but the internal twists dictate that it is Theory IIa\footnote{The difference between Theory IIa and Theory IIb, as proposed in \cite{Distler:2020tub}, is captured by the difference in the global form of their flavour symmetry groups. There, we found that the flavour symmetry of Theory IIa is $Sp(3)_8/\mathbb{Z}_2\times Sp(2)_8\times Sp(2)_7$, whereas the  flavour symmetry of Theory IIb is $\bigl(Sp(3)_8 \times Sp(2)_8)\bigr)/\mathbb{Z}_2\times Sp(2)_7$ --- the quotient by the diagonal subgroup of the centers of $Sp(3)$ and $Sp(2)$. A method to find the global form of the global symmetry group of any fixture in class $S$ has been proposed in \cite{Bhardwaj:2021ojs}. Here we see that this rather subtle difference affects the conformal manifold of an $Sp(2)$ gauging of the theory in a significant way. In \eqref{IIBsp2gauged}, the flavour symmetry of the resulting (family of) SCFTs is $(Sp(3)_8)/\mathbb{Z}_2\times Sp(2)_7\times SU(2)_5$ and the theory has trivial 1-form symmetry, whereas in  \eqref{IIAgauged}, the flavour symmetry is $Sp(3)_8\times Sp(2)_7\times SU(2)_5$ and the theory has a $\mathbb{Z}_2$ 1-form symmetry (since all local operators, before the gauging, are invariant under the center of the $Sp(2)$ that we gauge). This is in accord with the general results of  \cite{Bhardwaj:2021pfz}.}  in \cite{Distler:2020tub}. The twist around the degenerating cycle does not take the value $\beta$ or $\beta^2$, and hence, the gauge theory fixture containing $\tilde{A}_1$ does not appear.

In another degeneration limit, where $[4,1^2]$ collides with $[2,1^4]$, we obtain a $Spin(8)$ gauge theory

\begin{displaymath}
 \includegraphics[width=204pt]{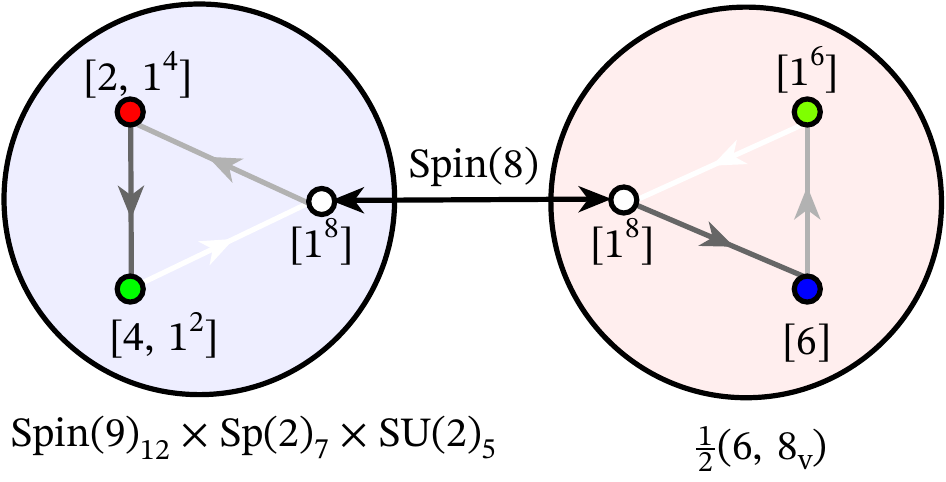}
\end{displaymath}
where the hypermultiplets transform in $\frac{1}{2}(6,8_v)$ instead of $\frac{1}{2}(6,8_s)$. Finally, in the third degeneration limit, we obtain a $Spin(7)$ gauge theory

\begin{displaymath}
 \includegraphics[width=204pt]{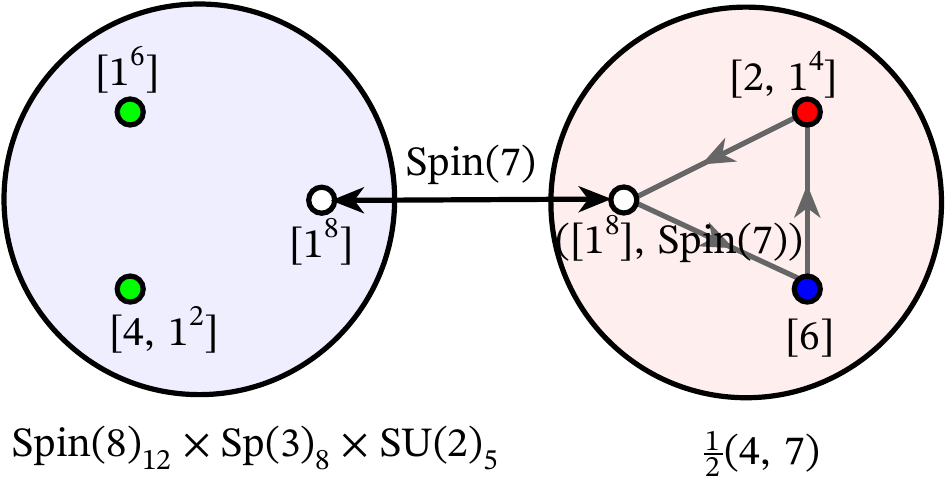}
\end{displaymath}
where the embedding of $Spin(7)$ in $Spin(8)$ differs from the one considered above by the action of triality.

The conformal manifold for this family of SCFTs is a copy of $\overline{\mathcal{M}}_{0,4}$. It is straightforward to check that we get the same family of SCFTs if we take $\delta_1 = \delta_2 = \delta_3 = \delta_4 = \alpha \beta^j$ for $j=1,2$ instead.

\subsection{Six weak coupling limits}\label{six_weak}

In \S\ref{resolving_atypical_punctures}, we saw that a 4-punctured sphere, with 4 punctures in the $[\alpha]$ conjugacy class could have as many as 6 distinct weak-coupling limits at various points on its 1-dimensional conformal manifold. But the example studied there had only 5 weak-coupling limits, as one degeneration of the surface yielded a gauge-theory fixture --- a strongly-coupled point in the interior of the conformal manifold. Here, we will exhibit an example where all of the degenerations are weakly-coupled and so we indeed get 6 distinct weak-coupled limits.

Let us start with a configuration, as before, with trivial internal twists

\begin{displaymath}
 \includegraphics[width=210pt]{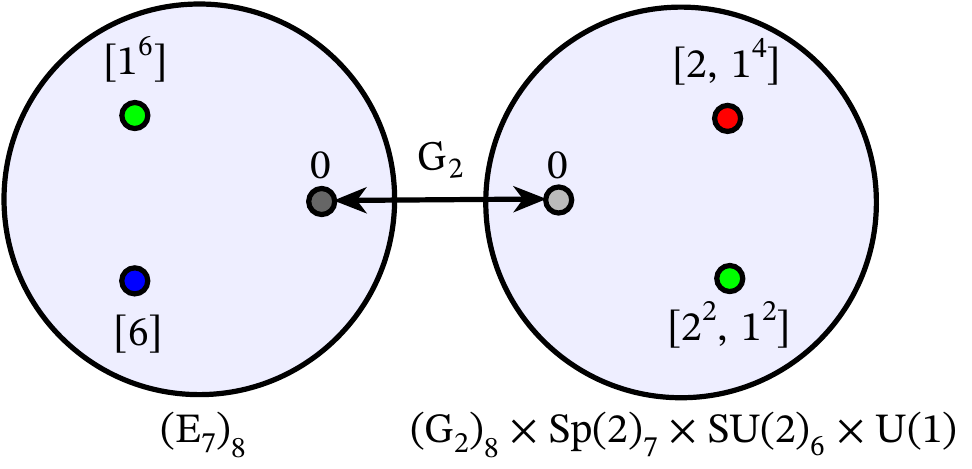}
\end{displaymath}
Over the same point on the boundary of $\overline{\mathcal{M}}_{0,4}$, we also find a $Spin(8)$ gauge theory

\begin{displaymath}
 \includegraphics[width=227pt]{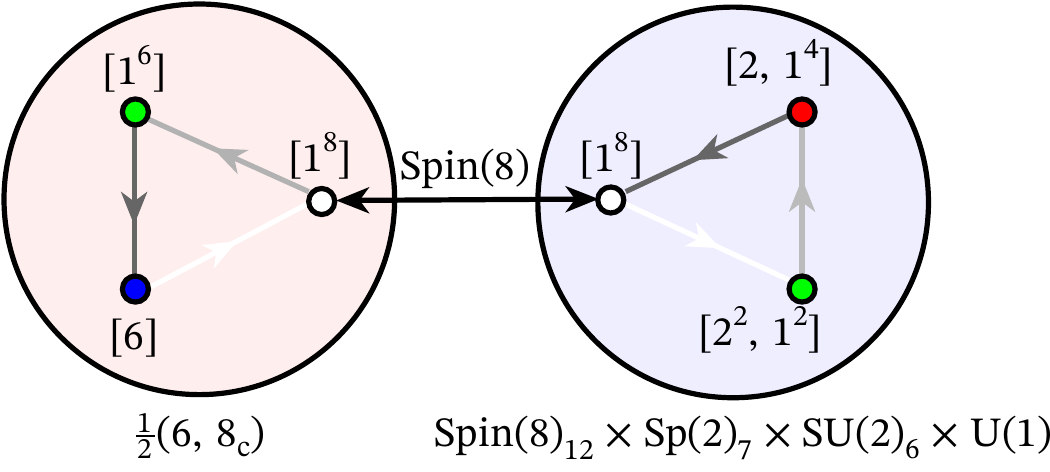}
\end{displaymath}
When $[1^6]$ collides with $[2,1^4]$ puncture, we again have two possible weak-coupling limits. When the twist around the degenerating cycle is $\mathbb{1}$, we get an $SU(4)$ gauge theory,

\begin{displaymath}
 \includegraphics[width=216pt]{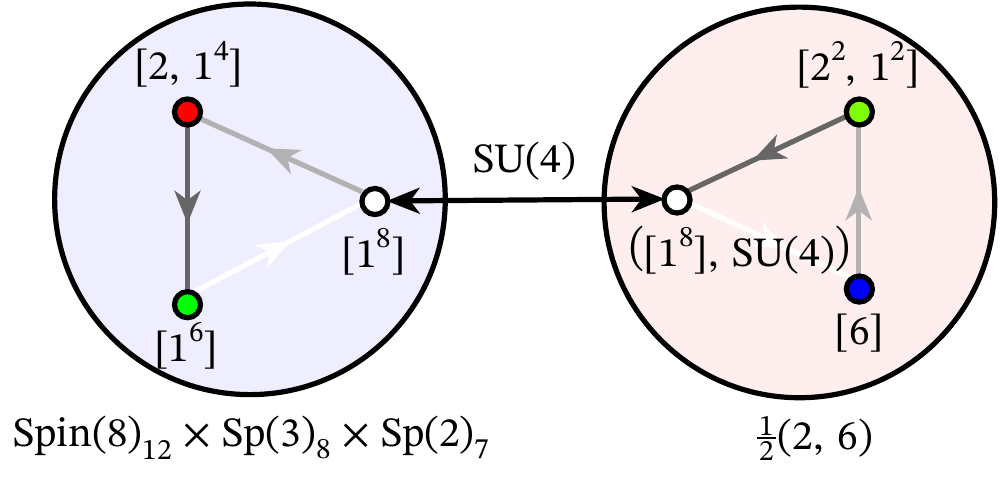}
\end{displaymath}
and when it is $\beta$ or $\beta^2$, we obtain an $SU(3)$ gauge theory.

\begin{displaymath}
 \includegraphics[width=194pt]{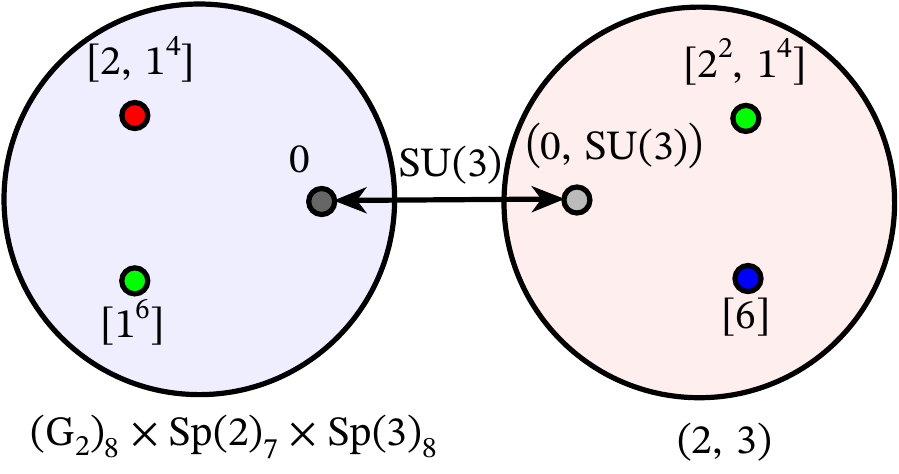}
\end{displaymath}
Finally, when $[1^6]$ collides with the $[2^2,1^2]$, we obtain a $Spin(7)$ gauge theory when the twist around the pinching cycle is $\mathbb{1}$,

\begin{displaymath}
 \includegraphics[width=204pt]{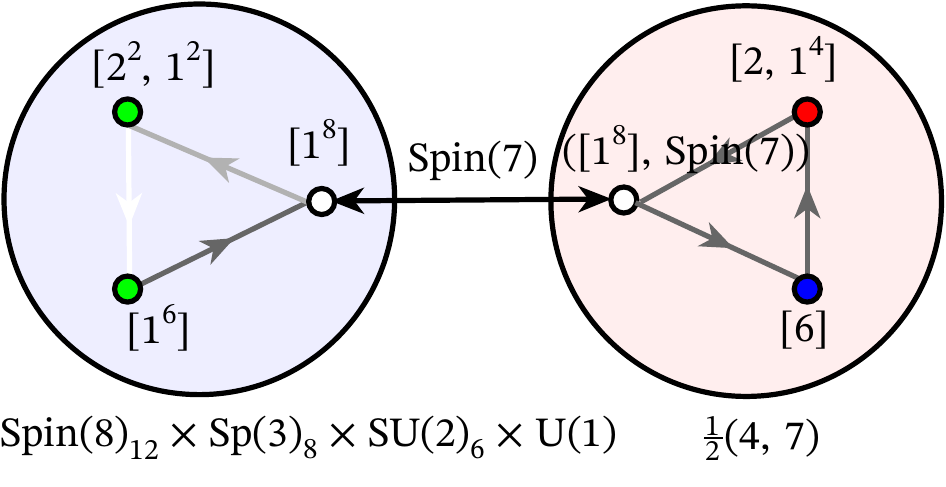}
\end{displaymath}
and when it is $\beta$ or $\beta^2$ we obtain a $G_2$ gauge theory.

\begin{displaymath}
 \includegraphics[width=190pt]{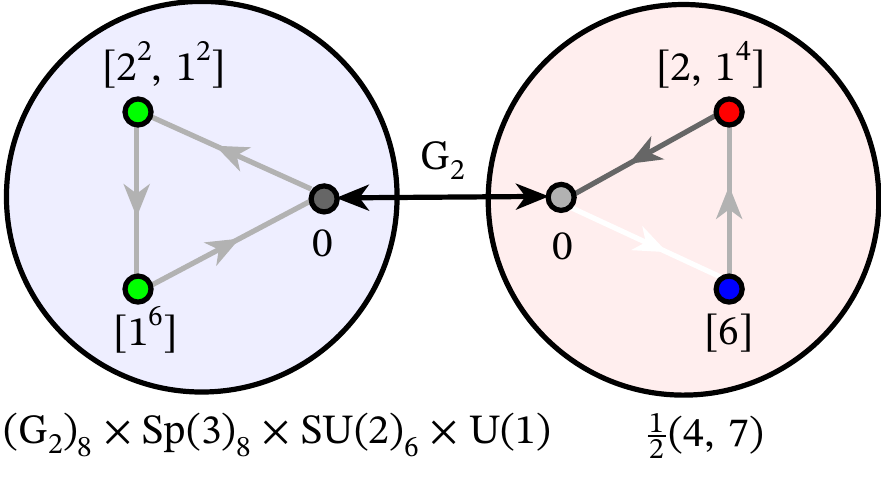}
\end{displaymath}

Here, the 6 weakly-coupled descriptions occur at the 6 cusp points of $\text{UHP}/\Gamma(4)$, a structure radically-different from what has been seen, heretofore, in $\mathcal{N}=2$ SCFTs. We will discuss what happens when we make other choices for the conjugacy classes of the external twists in \cite{Distler:toappear}.

\section*{Acknowledgements}
We would like thank Andrew Neitzke for useful discussions. This work was supported in part by the National Science Foundation under Grant No.~PHY--1914679. BE is supported in part
by the Israel Science Foundation under grant No. 1390/17. The work of AS was supported by NSF grant DMS-2005312. AS would also like to thank the Department of Mathematics at Yale University for their hospitality during completion of this work. 
\bibliographystyle{utphys}
\bibliography{ref}

\end{document}